\documentclass{ieeeaccess}
\usepackage{cite}
\usepackage{amsmath,amssymb,amsfonts}
\usepackage{graphicx,color}
\usepackage{textcomp}
\usepackage{hyperref}
\usepackage{algorithm}
\usepackage{algpseudocode}
\usepackage{placeins}
\usepackage{adjustbox}
\usepackage{amssymb}
\usepackage{bm}
\usepackage{multirow}
\usepackage{threeparttable}
\usepackage{makecell}
\usepackage{bbding}
\newtheorem{definition}{Definition}
\usepackage{balance}
\usepackage{textcomp}

\makeatletter
\AtBeginDocument{\DeclareMathVersion{bold}
	\SetSymbolFont{operators}{bold}{T1}{times}{b}{n}
	\SetSymbolFont{NewLetters}{bold}{T1}{times}{b}{it}
	\SetMathAlphabet{\mathrm}{bold}{T1}{times}{b}{n}
	\SetMathAlphabet{\mathit}{bold}{T1}{times}{b}{it}
	\SetMathAlphabet{\mathbf}{bold}{T1}{times}{b}{n}
	\SetMathAlphabet{\mathtt}{bold}{OT1}{pcr}{b}{n}
	\SetSymbolFont{symbols}{bold}{OMS}{cmsy}{b}{n}
	\renewcommand\boldmath{\@nomath\boldmath\mathversion{bold}}}
\makeatother

\def\BibTeX{{\rm B\kern-.05em{\sc i\kern-.025em b}\kern-.08em
		T\kern-.1667em\lower.7ex\hbox{E}\kern-.125emX}}

\begin{document}
	\history{Date of publication xxxx 00, 0000, date of current version xxxx 00, 0000.}
	\doi{10.1109/ACCESS.2024.0429000}
	
	\title{FQsun: A Configurable Wave Function-Based Quantum Emulator for Power-Efficient Quantum Simulations}
	\author{
		\uppercase{TUAN HAI VU}\authorrefmark{1}, \IEEEmembership{Member, IEEE},
		\uppercase{VU TRUNG DUONG LE}\authorrefmark{1}, \IEEEmembership{Member, IEEE},
		\uppercase{HOAI LUAN PHAM}\authorrefmark{1}, \IEEEmembership{Member, IEEE},
		\uppercase{QUOC CHUONG NGUYEN}\authorrefmark{2},
		and \uppercase{YASUHIKO NAKASHIMA}\authorrefmark{1}, \IEEEmembership{Senior Member, IEEE}}
	
	\address[1]{Nara Institute of Science and Technology, 8916–5 Takayama-cho, Ikoma, Nara 630-0192, Japan}
	\address[2]{Department of Mathematics, University at Buffalo, New York, NY 14260}
	
	\tfootnote{This work was supported by JST-ALCA-Next Program Grant Number JPMJAN23F4, Japan, and partly executed in response to the support of JSPS, KAKENHI Grant No. 22H00515, Japan.}
	
	\markboth
	{Tuan Hai Vu et al \headeretal: Preparation of Papers for IEEE TRANSACTIONS and JOURNALS}
	{Tuan Hai Vu et al \headeretal: Preparation of Papers for IEEE TRANSACTIONS and JOURNALS}
	
	\corresp{Corresponding author: VU TRUNG DUONG LE (e-mail: le.duong@naist.ac.jp).}

	\begin{abstract}
		Quantum computers are promising powerful computers for solving complex problems, but access to real quantum hardware remains limited due to high costs. Although the software simulators on CPUs/GPUs such as Qiskit, ProjectQ, and Qsun offer flexibility and support for many qubits, they struggle with high power consumption and limited processing speed, especially as qubit counts scale. Accordingly, quantum emulators implemented on dedicated hardware, such as FPGAs and analog circuits, offer a promising path for addressing energy efficiency concerns. However, existing studies on hardware-based emulators still face challenges in terms of limited flexibility and lack of fidelity evaluation. To overcome these gaps, we propose FQsun, a quantum emulator that enhances performance by integrating four key innovations: efficient memory organization, a configurable Quantum Gate Unit (QGU), optimized scheduling, and multiple number precisions. Five FQsun versions with different number precisions are implemented on the Xilinx ZCU102, consuming a maximum power of 2.41W. Experimental results demonstrate high fidelity, low mean square error, and high normalized gate speed, particularly with 32-bit versions, establishing FQsun’s capability as a precise quantum emulator. Benchmarking on famous quantum algorithms reveals that FQsun achieves a superior power-delay product, outperforming software simulators on CPUs in the processing speed range. 
	\end{abstract}
	
	
	\titlepgskip=-21pt
	
	\maketitle
	
	\section{Introduction}

	\PARstart{Q}{uantum} computing is a captivating research field that has many promising applications in factorial problem \cite{doi:10.1137/S0036144598347011}, searching \cite{grover1996fast}, optimization \cite{PhysRevApplied.19.024027}, or quantum machine learning \cite{Guan_2021}. Numerous researchers have developed real quantum devices successfully, such as IBM Quantum \cite{Bravyi2024}, Google Quantum AI \cite{arute2019quantum}, and QuEra \cite{Bluvstein2024}; some even claim to have achieved “quantum supremacy”. Recently, quantum computers have rapidly transformed from the Noise-Intermediate Scale Quantum (NISQ) era to the Fault-Tolerant Quantum Computer (FTQC) era, in which the logical error rate from quantum calculations is acceptable; it pursued the development of research in quantum computing. However, accessible real quantum computers such as IBM computers are limited because of the high cost. 
	
	To satisfy the growing demand for quantum computing simulations, several substantial research efforts have been conducted, as summarized in detail in Table~\ref{tab:related_work}. Key criteria include physical hardware implementation, benchmarking tasks, the number of simulated qubits (\#Qubits), precision, and evaluation metrics. Among existing quantum simulation software development kits (SDKs), the most well-known include Qiskit \cite{javadiabhari2024quantumcomputingqiskit}, ProjectQ \cite{Steiger2018projectqopensource}, Cirq \cite{isakov2021simulations}, and TensorFlow Quantum (TFQ) \cite{broughton2021tensorflowquantumsoftwareframework}, which are developed in Python. These SDKs not only support a wide range of quantum simulation models with high \#Qubits but also enable deployment on actual Quantum Processing Units (QPUs). Additional packages, such as QUBO \cite{date2021qubo}, PennyLane  \cite{bergholm2022pennylaneautomaticdifferentiationhybrid}, and cuQuantum \cite{10313722}, are also widely used due to their efficient performance on general-purpose processors, including Central Processing Units (CPUs) and Graphics Processing Units (GPUs). Notably, cuQuantum is optimized specifically for NVIDIA GPUs. Another notable package is Qsun, a quantum simulation package proposed in \cite{nguyen2022qsun}, which introduces a Wave Function (WF) - based quantum simulation system to reduce the computational load. Overall, these quantum simulation packages exhibit flexible and increasingly fast performance on software platforms. However, as the demand for higher \#Qubits grows, the reliance on 32- and 64-bit floating-point (FP) types and the general-purpose design of these software-based quantum simulators result in decreased processing speed and significant storage requirements. Consequently, these platforms tend to consume considerable energy without meeting speed requirements, an issue that is often overlooked in current research. \textit{In the future, if quantum simulations become widespread, simulators running on CPUs/GPUs, which consume between 150-350 W, will have a significant environmental impact.}
	
	To solve the limitations of software, various quantum emulation systems have been developed for hardware platforms, particularly analog circuits and Field-Programmable Gate Arrays (FPGA). Specifically, analog-based emulators, introduced in \cite{10265215}, use CMOS analog circuits to represent real numbers naturally, which are suitable for simulating quantum states. Although analog circuits achieve higher energy efficiency, scalability, and simulation speed than software, they face limitations when they only support a single algorithm like Grover's Search Algorithm (GSA) or Quantum Fourier Transform (QFT) with lower fidelity due to noise and other unintended effects. In contrast, FPGA-based quantum emulators, introduced in \cite{10653682, 10050794, Mahmud2020EfficientCT, 9952408, 10.1109/ISVLSI.2008.43, 1347938, https://doi.org/10.1155/2016/5718124, 8115369, 9798809, suzuki2022quantum }, offer a more practical and developable approach while still achieving high speed and energy efficiency. However, these studies lack detailed fidelity evaluations, particularly when using $8$-bit and $16$-bit fixed-point (FX) representations, which may not meet precision requirements. FPGA-based emulators are also limited by high storage and communication demands, restricting them to lower $\#$Qubits. Moreover, these FPGA emulators have low flexibility, as they are also optimized only for specific applications such as QFT, GSA, Quantum Haar Transform (QHT), and Quantum Support Vector Machine (QSVM). In general, existing related works continue to face specific challenges, including high energy consumption in quantum simulators, while hardware-based quantum emulators remain limited by low flexibility, reduced precision, and limited applicability due to implementation difficulties. \textit{Although they attempt to improve performance, most are constrained by the $\#$Qubit they can simulate and lack detailed evaluations of parameters that reflect accuracies, such as fidelity or mean-square error (MSE). This severely impacts their quality and practicality.}
	
	To address the aforementioned challenges, the FQsun quantum emulator is proposed to achieve optimal speed and energy efficiency. FQsun implements four innovative propositions: efficient memory organization, a configurable Quantum Gate Unit (QGU), optimal working scheduling, and multiple number precisions to maximize the flexibility, speed, and energy efficiency of the quantum emulator. FQsun is implemented on a Xilinx ZCU102 to demonstrate its effectiveness at the real-time SoC level. Through strict evaluation of execution time, accuracy, power consumption, and power-delay product (PDP), FQsun will demonstrate its superiority over software quantum simulators running on powerful Intel CPUs and NVIDIA GPUs, as well as existing hardware-based quantum emulators, when performing various quantum simulation tasks.

	\begin{table*}[]
		\caption{Surveyed results of related quantum simulators/emulators in the last five years. ($\dagger$) Estimated qubit numbers based on projected performance given hardware constraints.}
		\label{tab:related_work}
		\renewcommand{\arraystretch}{1.2}
		\centering
		\small
		\begin{tabular}{|p{50pt}|p{70pt}|p{90pt}|p{60pt}|p{60pt}|p{100pt}|}
			\hline
			\textbf{Reference} & \textbf{Devices} & \textbf{Benchmarking tasks} & \textbf{\#Qubits} & \textbf{Precision} & \textbf{Comparison metrics} \\ \hline
			Qiskit \cite{javadiabhari2024quantumcomputingqiskit} & CPUs, GPUs, QPUs & QFT, BV, Random Clifford circuits, Hamiltonian simulation, QAOA & Depending on hardware resources & 32-bit FP, 64-bit FP & Execution time, Gate count, Gate depth, Fidelity \\ \hline
			ProjectQ \cite{Steiger2018projectqopensource} & CPUs, GPUs, QPUs & QFT, Shor's Algorithm & Depending on hardware resources & 32-bit FP, 64-bit FP & Execution time, Resource utilization, Circuit depth, Gate count, Fidelity \\ \hline
			Cirq \cite{isakov2021simulations} & CPUs, GPUs, QPUs & Random Quantum Circuits (RQC) & Depending on hardware resources & 32-bit FP & Execution time, Fidelity \\ \hline
			TFQ \cite{broughton2021tensorflowquantumsoftwareframework} & CPUs, GPUs, QPUs & QAOA, QNN & Depending on hardware resources & 32-bit FP, 64-bit FP & Execution time, Fidelity, Gradient-based optimization \\ \hline
			QUBO \cite{date2021qubo} & CPUs, GPUs & QFT, VQE, GSA, Adiabatic Time Evolution, Quantum Autoencoder, Quantum Classifier & Depending on hardware resources & 32-bit FP, 64-bit FP & Execution time, Fidelity, Overlap \\ \hline
			PennyLane \cite{bergholm2022pennylaneautomaticdifferentiationhybrid} & CPUs, GPUs & VQE, QAOA, Quantum Classification & Depending on hardware resources & 32-bit FP, 64-bit FP & Execution time, Fidelity, Variational accuracy \\ \hline
			cuQuantum SDK \cite{10313722} & GPUs & QFT, QAOA, Quantum Volume, Phase Estimation & Depending on hardware resources & 32-bit FP, 64-bit FP & Execution time, Speedup factor, Memory bandwidth utilization \\ \hline
			Qsun \cite{nguyen2022qsun} & CPUs & QLR, QNN, QDP & Depending on hardware resources & 32-bit FP, 64-bit FP & Execution time, Fidelity, MSE \\ \hline
			Analog-based emulator \cite{10265215} & UMC-180 nm CMOS Analog & GSA, QFT & 6 to 17 qubits & 32-bit FP & Execution time, Power consumption, Circuit compatibility \\ \hline
			ZCVU9P-based emulator \cite{10653682} & FPGA(Xilinx ZCVU9P) + CPU(Xeon E5-2686 v4) & QSVM, Quantum Kernel Estimation & Up to 6 qubits & 16-bit FX & Execution time, Numerical accuracy, Test accuracy \\ \hline
			Alveo-based emulator \cite{10050794} & FPGA (Xilinx Alveo U250) & QHT, 3D-QHT & 10 to 32 qubits ($\dagger$) & 32-bit FP & Execution time, Resource utilization, Circuit depth \\ \hline
			Arria-based emulator \cite{Mahmud2020EfficientCT} & FPGA (Arria 10AX115N4F45E3SG) & GSA & Up to 32 qubits ($\dagger$) & 16-bit FX & Emulation time, Resource utilization, Frequency \\ \hline
			XCKU-based emulator \cite{9952408} & FPGA(Xilinx XCKU115) & QFT & Up to 16 qubits & 16-bit FX & Execution time \\ \hline
			Stratix-based emulator \cite{10.1109/ISVLSI.2008.43, 1347938} & FPGA(Altera Stratix EP1S80B956C6) & QFT, GSA & Up to 8 qubits & 8-bit FX, 16-bit FX & Logic cell usage, Emulation time, Frequency \\ \hline
			\textbf{This work FQsun} & \textbf{FPGA(Xilinx ZCU102)} & \textbf{QFT, PSR (QDP), RQC, QLR, QNN, etc.} & \textbf{3 to 17 qubits} & \textbf{16-bit FX/FP, 24-bit FX, 32-bit FX/FP} & \textbf{Execution time, PDP, Fidelity, MSE, Power, Hardware Utilization} \\ \hline
		\end{tabular}
	\end{table*}

	\begin{figure*}[t!]
		\centering
		\includegraphics[width=0.95\linewidth]{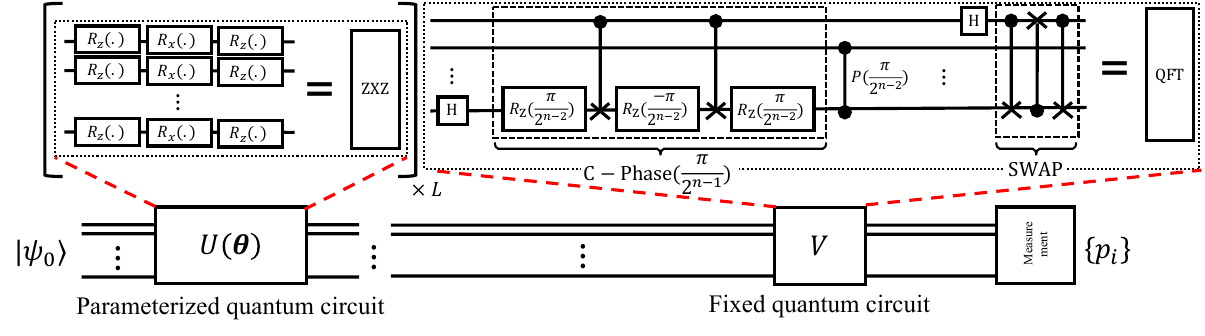}
		\caption{Quantum circuit simulation model where it can divide into parameterized and fixed parts, denoted as $U(\bm{\theta})$ and $V$, respectively (both can be notated as $U^{(\bm{t})}$). (Inset left) An example of a parameterized part is the ZXZ layer. (Inset right) An example of a fixed part is the QFT circuit; each part inside consists of many basic gates.}
		\label{fig:overview}
	\end{figure*}
	
	The outline of this paper is organized as follows. Section \ref{sec:background} presents the background of the study. Next, details of the FQsun hardware architecture are presented in Section \ref{sec:proposal}. Then, benchmarking results, evaluation, and comparisons are elaborated upon in Section \ref{sec:result}. Finally, Section \ref{sec:conclusion} concludes the paper.

	\section{BACKGROUNDS}
	\label{sec:background}
	
	\subsection{QUANTUM SIMULATOR}
	
	Due to the wide range of applications and programming languages, many quantum simulators have been proposed, from low-level languages such as C to high-level languages such as Python. The goals of quantum simulator development range from domain-specific to general purpose, then reach the boundary of quantum advantage where classical computers can no longer simulate a quantum system in acceptable runtime. The basic operation in $n$ - qubit system can be expressed as $|\psi^{(\bm{m})}\rangle = \mathcal{U}(\bm{\theta})|\psi^{(\bm{0})}\rangle$ parameterized by $\bm{\theta}$, where $\mathcal{U}(\bm{\theta})$ is composed from $\{g_j\}_{j=1}^{m} $ and $m$ is the number of quantum gates (gates). 
	$\mathcal{U}(\bm{\theta})$ can be presented as $\bigotimes_{j=1} U^{(\bm{t})}$ which is the tensor product between operators.
	
	The quantum state $|\psi^{(\bm{t})}\rangle$ is a column complex vector $[\alpha_0^{(\bm{t})}\; \alpha_1^{(\bm{t})}\; \ldots\; \alpha_{N-1}^{(\bm{t})}]^{\intercal}= \sum_{j=0}^{N-1} \alpha_j^{(\bm{t})}|j\rangle$ with $|j\rangle$ are elements of the computational basis and $\{\alpha_j^{(\bm{t})}\}$ are complex entries ($N=2^n$). $\mathcal{U}(\bm{\theta})$ and $|\psi^{(\bm{t})}\rangle$ satisfy unitary and normalize condition, i.e, 
	$(U^{(\bm{t})})^{\dagger}U^{(\bm{t})}=\mathbb{I}$ and $\;\langle\psi^{(\bm{t})}|\psi^{(\bm{t})}\rangle =1$.
	
	The target of simulating quantum computing is to get the expectation value when measuring a final quantum state under an observable $\hat{B}:\langle E \rangle = \langle \psi | \hat{B} |\psi\rangle$, which is known as the physical property of the system, summarized as Definition~\ref{def:simulation}. 
	
	\begin{definition}[Strong and weak simulation] Given a quantum circuit $U$ and any specified object $x$, the strong simulation can compute any value $\langle \bm{0} | U | x \rangle$ and a weak simulation  samples from the distribution $p(x)=|\langle \bm{0} | U | x \rangle|^2$ \cite{9123426, 10.1007/978-3-031-65633-0_25}.
		\label{def:simulation}
	\end{definition}
	
	The term $\mathcal{U}(\bm{\theta})|\psi^{(\bm{0})}\rangle$ can be computed in various approaches mainly by Matrix Multiplication (MM) \cite{10.1145/3183895.3183901, isakov2021simulationsquantumcircuitsapproximate, 10313722}, Tensor-network \cite{10313711}, Near-Clifford \cite{Descamps_2024, gidney2021stim} or WF. In principle, MM conducts vanilla matrix-vector multiplication between operators and state vector, consuming $\mathcal{O}(m \times 2^{2n})$ in time complexity and $\mathcal{O}(2^{2n}+2^n)$ in space complexity for storing both operator and state vector. MM can be accelerated through some strategies, including gate fusion \cite{10.1145/3126908.3126947}, indexing, realized-state representation \cite{https://doi.org/10.4218/etrij.2021-0442}, or sparse matrix calculation. The other approach represents quantum circuits using their corresponding WF \cite{Steiger2018projectqopensource, Suzuki2021qulacsfast}. \textit{The WF approach, in theory, requires only $\mathcal{O}(2^n)$ in space complexity for storing only the state vector, which has fewer computational operations and less temporary memory space for conducting quantum operations.}
	
	\subsection{SIMULATING QUANTUM CIRCUIT BY WAVE-FUNCTION APPROACH}
	
	The MM approach operates in two steps. Firstly, $U^{(\bm{t})}$ is constructed by tensor-product between $g_j$ and $\mathbb{I}$; later, we conduct MM operation between $U^{(\bm{t})}$ and $|\psi^{(\bm{t-1})}\rangle$, where $U^{(\bm{t})}|\psi^{(\bm{t-1})}\rangle=|\psi^{(\bm{t})}\rangle$.
	With the unfavorable, $U^{(\bm{t})}\in\mathbb{C}^{N \times N}$, there is an exponential resource scaling of both state vector and operators with increasing $\#\text{Qubits}$. 
	
	Instead of constructing a large matrix such as $U^{(\bm{t})}$, The WF approach uses $m$ gates $\{g_j\}$ directly, took only $2^{n + \hat{n}_j + 1}$ multiplication and $2^{n + \hat{n}_j + 1}$ addition per gate, with $\hat{n}_j\in[1,2]$ is the number of operand of the gate $j$; instead of $2^{2n}$ multiplication and $2^{2n-2}$ addition following by MM approach. The idea behind this technique is to update only non-zero amplitudes in $|\psi^{(\bm{t})}\rangle$ sequentially:
	
	\begin{equation}
		|\psi^{(\bm{m})}\rangle=\mathcal{W}(g_m)\ldots \mathcal{W}(g_2)\mathcal{W}(g_1)|\psi^{(\bm{0})}\rangle,
		\label{eq:wave-function}
	\end{equation}
	where $g\in \mathcal{G}=\{H,S,CX,R_i(.)\},i\in \{x,y,z\}$, $\{H,S\}$ and $CX$ are constant $2\times 2$ and $4\times 4$ matrices, respectively. $R_i(.)$ is the $2\times 2$ parameterized matrix with single parameter $\theta\in[0,2\pi]$. $\text{Clifford}+R_i$ is chosen for two purposes. First, the universal gate set known as Clifford + T belongs to the above set as Definition~\ref{def:gate_set}, where the $T$ gate is equivalent to the $R_z(\pi/4)$ gate, and the second is that simulating the variational quantum model requires the use of parameterized gates.
	
	\begin{definition}[Gate set for strong simulation] Clifford group include $\{H, S, CX\}$ can be used to simulate by the classical algorithm in polynomial time and space, follow Gottesman-Knill theorem \cite{PhysRevA.70.052328}, which can be extended to Clifford + $R_i$.
		\label{def:gate_set}
	\end{definition}
	
	The transition function $\mathcal{W}(g)$ updates amplitudes $\{\alpha_j\}$ directly depending on the type of gate \cite{Jones2019}, with the acting on the $(w_0)^{\text{th}}$ qubit as Equation~\eqref{eq:gate_action}:
	
	\begin{equation}
		\mathcal{W}(g):\begin{bmatrix} \alpha_{s_i} \\ \alpha_{s_i+2^{w_0}} \end{bmatrix} \rightarrow g \begin{bmatrix} \alpha_{s_i} \\ \alpha_{s_i+2^{w_0}}\end{bmatrix},  
		\label{eq:gate_action}
	\end{equation}
	
	where $s_i=\lfloor i / 2^{w_0}\rfloor 2^{w_0+1}+\left(i \bmod 2^{w_0}\right)$, for all $i \in\left[0,2^{n-1}-1\right]$. We unify the implementation of single- and multiple-qubit gates into a common framework; the operation of single-qubit and $CX$ gates are outlined in Algorithm.~\ref{algo:basic} and  Algorithm.~\ref{algo:basic2}, respectively.

	\begin{algorithm}
		\caption{Operation of a single-qubit gate}
		\label{algo:basic}
		\begin{algorithmic}[]
			\Require $\{\alpha^{(\bm{t})}_j\}$, target qubit $w_0\in[0,n)$, $g = \begin{bmatrix}
				a & b\\
				c & d
			\end{bmatrix}\in \mathcal{G}$
			\Ensure $\sum_j|\alpha^{(\bm{t})}_j|^2=1$
			\State $n \gets \log_2 |\{\alpha_j\}|, \text{cut} \gets 2^{n-w_0-1}$
			
			\State $|\psi^{(\bm{t+1})}\rangle \gets[\alpha^{(\bm{t+1})}_0\;\alpha^{(\bm{t+1})}_1\;\ldots\;\alpha^{(\bm{t+1})}_{N-1}]^{\intercal}= [0\;0\;\ldots\;0]^{\intercal},$
			
			\State $\text{state}\gets[\text{bin}(0, n), \text{bin}(1, n),\ldots,\text{bin}(N-1, n)]$ \Comment{$\text{bin}(i,k)$ is the function that represent $i$ by $k$ binary bit},
			\For{$i \gets [0,1,\ldots, N-1]$}
			\If{$\text{state}[i][w_0] == 0$}
			\State $\alpha^{(\bm{t+1})}_i \gets \alpha^{(\bm{t+1})}_i + a \times \alpha^{(\bm{t})}_i$
			\State $\alpha^{(\bm{t+1})}_{i+\text{cut}} \gets \alpha^{(\bm{t+1})}_{i + \text{cut}} + b \times \alpha^{(\bm{t})}_i$
			\Else
			\State $\alpha^{(\bm{t+1})}_i \gets \alpha^{(\bm{t+1})}_i + d \times \alpha^{(\bm{t})}_i$
			\State $\alpha^{(\bm{t+1})}_{i-\text{cut}} \gets \alpha^{(\bm{t+1})}_{i-\text{cut}} + c\times \alpha^{(\bm{t})}_i$
			\EndIf
			\EndFor
			\State \Return $\{\alpha^{(\bm{t+1})}_j\}$
		\end{algorithmic}
	\end{algorithm}

	\subsection{PRELIMINARY CHALLENGES FOR DEVELOPING A HIGH-EFFICIENCY QUANTUM EMULATOR}
	\label{sec:challenge}

	To develop a quantum emulator that meets the demands of speed, flexibility, and energy efficiency, several core challenges must be addressed. Each of these challenges highlights essential requirements that drive the need for innovative design improvements in quantum emulation systems:
	
	\textbf{Challenge 1: Efficient Memory Management}. Quantum emulation requires managing and processing large datasets, especially as the number of qubits increases, which dramatically expands memory demands. Efficient data handling and minimized memory access times are crucial to achieving high processing speeds without excessive energy consumption.
	
	\textbf{Challenge 2: Flexible Quantum Gate Configuration}. Different quantum algorithms rely on a variety of quantum gates, but many emulators are constrained by limited gate support. To effectively execute diverse algorithms, an adaptable gate configuration system is needed to facilitate a wide range of computations with minimal hardware reconfiguration.
	
	\textbf{Challenge 3: Optimized Task Scheduling}. Quantum emulators must execute sequential calculations, where delays in one step can impact the entire system's performance. To avoid bottlenecks, the emulator requires an optimized scheduling mechanism to ensure continuous processing without latency, thus maximizing resource utilization and computation speed.
	
	\textbf{Challenge 4: Multi-Level Precision Support}. Quantum algorithms often demand different levels of numerical precision, from FX to FP, depending on application requirements. Supporting various precision levels allows the emulator to balance accuracy with resource usage, adapting performance to meet the specific needs of each quantum task while minimizing computational overhead.
	
	Each of these challenges underscores the necessity of a robust, scalable emulator that can dynamically adjust to complex quantum workloads while maintaining efficiency and precision across diverse applications.
	
	\begin{algorithm}[t]
		\caption{Operation of CX gate} 
		\label{algo:basic2}
		\begin{algorithmic}[]
			\Require $\{\alpha^{(\bm{t})}_j\}$, control-target qubits $w_0,w_1\in[0,n)$, $g = \begin{bmatrix}
				\mathbb{I} & \bm{0}\\
				\bm{0} & \mathbb{X}
			\end{bmatrix}\in \mathcal{G}$
			\Ensure $\sum_j|\alpha^{(\bm{t})}_j|^2=1$
			\State $n \gets \log_2 |\{\alpha_j\}|, \text{cut} \gets 2^{n-w_1-1}$
			\State $|\psi^{(\bm{t+1})}\rangle \gets [\alpha^{(\bm{t+1})}_0\;\alpha^{(\bm{t+1})}_1\;\ldots\;\alpha^{(\bm{t+1})}_{N-1}]^{\intercal} = [0\;0\;\ldots\; 0]^{\intercal},$
			\State $\text{state} \gets [\text{bin}(0, n), \text{bin}(1, n),\ldots,\text{bin}(N-1, n)],$
			\For{$i \gets [0,1,\ldots, N-1]$}
			\If{$\text{state}[i][w_0] == 1$}
			\If{$\text{state}[i][w_1] == 1$}
			\State $\alpha^{(\bm{t+1})}_{i-\text{cut}} \gets \alpha^{(\bm{t+1})}_{i-\text{cut}} + \alpha^{(\bm{t})}_i$
			\Else
			\State $\alpha^{(\bm{t+1})}_{i+\text{cut}} \gets \alpha^{(\bm{t+1})}_{i+\text{cut}} + \alpha^{(\bm{t})}_i$
			\EndIf
			\Else
			\State $\alpha^{(\bm{t+1})}_{i} \gets \alpha^{(\bm{t})}_{i}$
			\EndIf
			\EndFor
			\State \Return $\{\alpha^{(\bm{t+1})}_j\}$
		\end{algorithmic}
	\end{algorithm}

	\section{PROPOSED ARCHITECTURE}
	\label{sec:proposal}

	\subsection{SYSTEM OVERVIEW ARCHITECTURE}

	\begin{table}[]
		\caption{Supported gates on FQsun, including Clifford + $R_i$ set with additional gates $\{T,X,Y,Z\}$.}
		\label{tab:gates}
		\centering
		\renewcommand{\arraystretch}{1.2}
		\begin{tabular}{|p{20pt}|p{95pt}|p{30pt}|p{40pt}|}
			\hline
			\textbf{Gate} & \textbf{Matrix Representation} & \textbf{Param} & \textbf{Equivalent to}\\ \hline
			
			$H$ & $\frac{1}{\sqrt{2}}\begin{bmatrix}1 & 1 \\ 1 & -1\end{bmatrix}$ & $w_0$ & Basic gate\\ \hline
			
			$S$ & $\begin{bmatrix}1 & 0 \\ 0 & i\end{bmatrix}$ & $w_0$ & Basic gate\\ \hline
			
			$CX$ & $\begin{bmatrix}1 & 0 & 0 & 0 \\ 0 & 1 & 0 & 0 \\ 0 & 0 & 0 & 1 \\ 0 & 0 & 1 & 0\end{bmatrix}$ & $w_0,w_1$ & Basic gate\\ \hline
			
			$R_x(\theta)$ & $\begin{bmatrix}\cos \left(\frac{\theta}{2}\right) & -i \sin \left(\frac{\theta}{2}\right) \\ -i \sin \left(\frac{\theta}{2}\right) & \cos \left(\frac{\theta}{2}\right)\end{bmatrix}$ & $w_0, \theta$ & Basic gate \\ \hline
			
			$R_y(\theta)$ & $\begin{bmatrix}\cos \left(\frac{\theta}{2}\right) & -\sin \left(\frac{\theta}{2}\right) \\ \sin \left(\frac{\theta}{2}\right) & \cos \left(\frac{\theta}{2}\right)\end{bmatrix}$ & $w_0, \theta$ & Basic gate\\ \hline
			
			$R_z(\theta)$ & $\begin{bmatrix}e^{-i \frac{\theta}{2}} & 0 \\ 0 & e^{i \frac{\theta}{2}}\end{bmatrix}$ & $w_0, \theta$ & Basic gate\\ \hline
			
			$T$ & $\begin{bmatrix} 1 & 0 \\ 0 & e^{-i\pi/4}\end{bmatrix}$ & $w_0$ & $R_z(\pi/4)$ \\ \hline
			
			$X$ & $\begin{bmatrix} 0 & 1 \\ 1 & 0\end{bmatrix}$ & $w_0$ & $R_x(\pi)$ \\ \hline
			
			$Y$ & $\begin{bmatrix} 0 & -i \\ i & 0\end{bmatrix}$ & $w_0$ & $R_y(\pi)$ \\ \hline
			
			$Z$ & $\begin{bmatrix} 1 & 0 \\ 0 & -1\end{bmatrix}$ & $w_0$ & $R_z(\pi)$ \\ \hline
			
		\end{tabular}
	\end{table}

\begin{figure}
		\centering
		\includegraphics[width=0.99\linewidth]{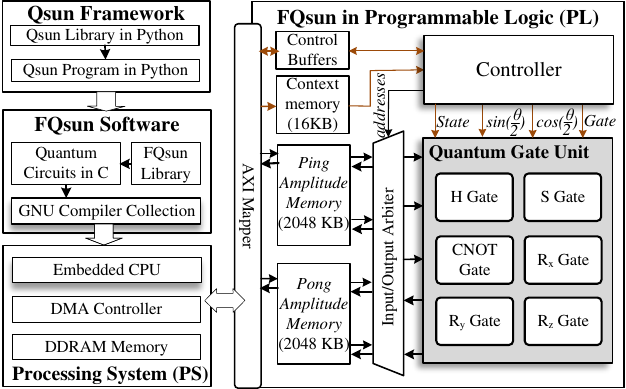}
		\caption{Overview architecture of our FQsun at the system-on-chip (Soc) level on FPGA.}
		\label{fig:soc_overview}
	\end{figure}

	Figure \ref{fig:soc_overview} details the proposed FQsun architecture on an FPGA at the system-on-chip (SoC) level. The system comprises four components: Qsun Framework, FQsun Software, Processing System (PS), and FQsun in Programmable Logic (PL). 
	
	The Qsun Framework, implemented in Python, includes libraries and programs for quantum simulation. These programs can instruct the hardware to execute gates through the FQsun C library in FQsun Software. Users proficient in C programming can directly write C programs by calling quantum circuits defined in the FQsun library to achieve higher performance. Within the PS, the embedded CPU runs software such as the Qsun Framework and FQsun Software to control FQsun in PL for executing quantum simulation tasks. In the PL, the proposed FQsun consists of four modules: AXI Mapper, memory system, FQsun controller, and QGU. The AXI Mapper manages and routes data exchanged between the PS and the FQsun memory system. The FQsun memory system comprises Control Buffers, Context Memory, and a pair of Ping/Pong Amplitude Memory modules. Control Buffers store control signals for the host FQsun Controller. Context Memory holds configuration information such as gate type, gate location ($cut$, $\sin(\theta/2)$, $\cos(\theta/2)$, $w_0$, $w_1$), and gate parameters. Notably, the Ping/Pong Amplitude $2\times 2048$ (KB) Memory is the largest memory unit designed to store $\{\alpha^{(t)}_j\}$ for QGU. Finally, QGU serves as the main processing unit of FQsun. QGU contains six fundamental gate units: $\{H, S, CX, R_x, R_y, R_z \}$. These basic gates can be combined to implement most other gates, ensuring high flexibility. The system overview architecture is designed to ensure compatibility between the PS and FQsun in hardware for stable and optimal operation. By supporting the FQsun library in Python and C, users can easily utilize FQsun to perform quantum tasks.

	\subsection{EFFICIENT MEMORY ORGANIZATION}
	
\begin{figure}[t!]
		\centering
		\includegraphics[width=0.99\linewidth]{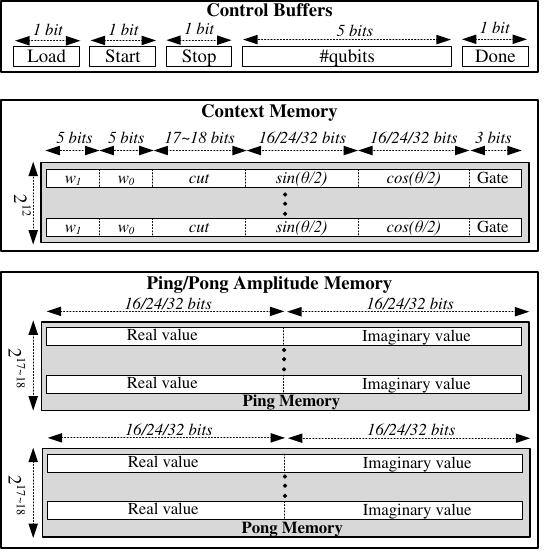}
		\caption{Memory organization of the proposed FQsun.}
		\label{fig:memory_organization}
		\vspace{-5mm}
	\end{figure}

	As mentioned in Section~\ref{sec:challenge}, matrix size $N$ grows exponentially with $\#\text{Qubits}$ that the emulator supports. Meanwhile, hardware platforms such as FPGAs are limited by storage resources. To support large $\#\text{Qubits}$, the memory organization of FQsun must be optimized. Figure \ref{fig:memory_organization} illustrates the memory organization of the proposed FQsun for efficient operation on FPGA. There are three types of memory: Control Buffers, Context Memory, and Ping/Pong Amplitude Memory.
	
	\textbf{Control Buffers:} These buffers store 1-bit control signals such as load, start, done, and stop. The load signal is used by the host to indicate the state where it writes context and amplitude data. Additionally, the host uses a 5-bit signal to indicate $\#\text{Qubits}$, which FQsun is supporting. After completing the load state, the host triggers the start signal to begin the session. The done signal is continuously read by the host to track the completion time of each gate operation. Finally, the stop signal is used to end the session by resetting the address register of the Context Memory.

	\textbf{Context Memory ($\bm{16}$ KB):} This memory stores configuration data for FQsun and has $d=2048$. The configuration data includes a $3$-bit gate opcode indicating the type of gate to be executed. The values $\sin(\theta/2)$ and $\cos(\theta/2)$, with a bit width of $16/24/32$ depending on the precision type, are parameters for $R_i$, pre-computed by FQsun software to reduce hardware complexity. Additionally, parameters such as the $5$-bit cut and gate positions $\{w_0, w_1\}$ must also be stored. With this Context Memory design, FQsun can support the execution of up to $2048$ gates in a computational sequence.
	
	\textbf{Ping/Pong Amplitude Memory ($\bm{2}$ MB):} This is the largest memory and is used to store $\{\alpha^{(\bm{t})}_j\}$ for QGU. This memory must have a $d=2^{17}$ or $d=2^{18}$ to support up to $17$ or $18$ qubits for $16$-bit precision or $32$-bit precision, calibrated to fit the maximum BRAM capacity of the Xilinx ZCU102. In case Ping/Pong Amplitude Memory is applied to FPGAs with more abundant BRAM resources, such as the Alveo or Versal series, FQsun could support a larger $\#\text{Qubits}$. To enable amplitude updates as described in Algorithms \ref{algo:basic} and \ref{algo:basic2}, a dual-port Ping/Pong memory design is applied. Specifically, if the ping memory holds the current amplitude $\{\alpha^{(\bm{t})}_j\}$, the pong memory will store the new amplitude $\{\alpha^{(\bm{t+1})}_j\}$, and vice versa. This design is essential on the grounds that, during the calculation of $\{\alpha^{(\bm{t+1})}_{i+cut}\}$ and $\{\alpha^{(\bm{t+1})}_{i-cut}\}$, both the current and new amplitudes are needed. The use of Ping/Pong memory allows for continuous amplitude updates, thereby optimizing simulation performance.
	
	\subsection{CONFIGURABLE QUANTUM GATE UNIT (QGU)}

	Existing quantum emulators face difficulties in supporting a wide variety of quantum simulation applications. This challenge arises due to the limited flexibility of their computational hardware architectures in executing complex gates critical for many applications, thereby restricting their capabilities. Consequently, the QGU hardware architecture is proposed to enable FQsun to achieve the highest level of flexibility in supporting any quantum simulation applications. Supported gates on FQsun, including Clifford + $R_i$ set with additional gates $\{T, X, Y, Z\}$.
	
	\begin{figure}[t!]
		\centering
		\includegraphics[width=0.99\linewidth]{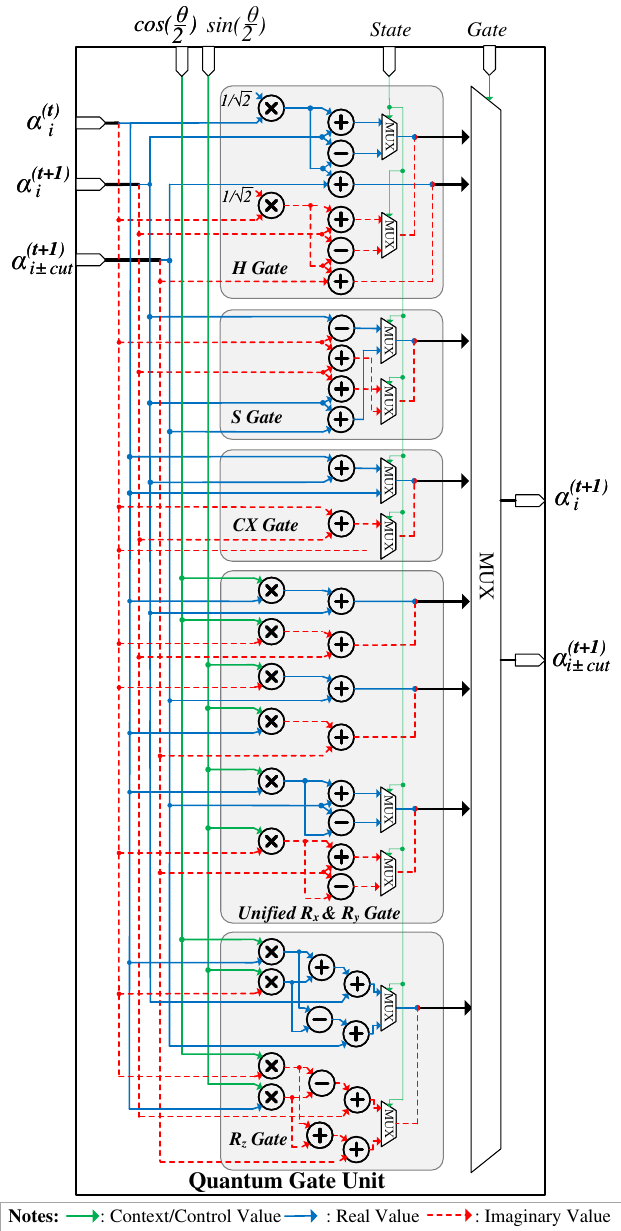}
		\caption{Micro-architecture of QGU.}
		\label{fig:qgu}
	\end{figure}
	
	The micro-architecture of the QGU, as detailed in Figure \ref{fig:qgu}, consists of five main modules designed to perform six fundamental gates following software design. The computational data for the QGU includes $\{\alpha^{(\bm{t})}_{i}\}$, $\{\alpha^{(\bm{t+1})}_{i}\}$, $\{\alpha^{(\bm{t+1})}_{i\pm cut}\}$, and parameters ($\sin(\theta/2)$ and $\cos(\theta/2)$) used by $R_i$. Essentially, the operations of $R_x$ and $R_y$ are relatively similar; hence, they are implemented using \textit{a Unified $R_x$ \& $R_y$ Gate unit} to conserve hardware resources. To ensure proper execution of the \verb+if-else+ logic as outlined in Algorithms. \ref{algo:basic} \& \ref{algo:basic2}, a 2-to-1 multiplexer system controlled by the state value is used to select the output value. Finally, a multiplexer controlled by the gate operation input is employed. By combining the fundamental gates to form other gates, the QGU can be configured to execute all gates utilized by Qsun\cite{nguyen2022qsun}, enabling diverse quantum applications. Moreover, to ensure the highest operating frequency, the operators within the quantum gate modules, such as multiplication and addition, are pipelined to shorten the critical path. Specifically, for FP numbers, multiplication and addition each require two stages, while for FX numbers, multiplication requires two stages, and addition does not require any stage.

	\subsection{OPTIMAL WORKING SCHEDULING}

	When operating at the SoC level, without a well-defined timing schedule for read and write operations in the memory system, FQsun cannot operate efficiently due to various issues, such as incorrect timing for control signal transmission when the computational data is not available or incorrect data read/write in the Ping/Pong Amplitude Memory.

\begin{figure}[t!]
		\centering
		\includegraphics[width=0.99\linewidth]{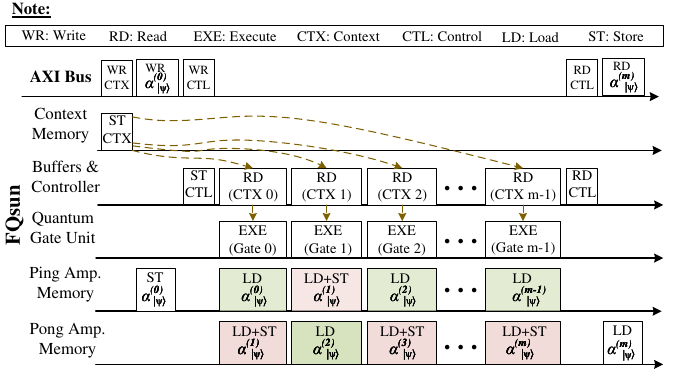}
		\caption{Timing chart of FQsun operation using Ping/Pong Amplitude Memories.}
		\label{fig:working_Flow}
	\end{figure}

	To address these issues, the timing diagram of FQsun for executing a working session is detailed in Figure \ref{fig:working_Flow}. At the beginning of a session, context data is stored in the Context Memory, which contains configuration commands for FQsun to execute a quantum circuit consisting of $m$ gates for a specific application. Next, the system writes initial amplitudes $\{\alpha^{(\bm{0})}_j\}$ into the Ping Amplitude Memory. The read/write process for amplitude data incurs the highest memory access time because the amplitude vector is large ($N = 2^n$). Once the ping memory is fully loaded with data, the system sends a control signal to initiate the FQsun working session. An internal control register called the Program Counter (PC) is used to track the instruction index to determine the current gate for execution. After reading the instruction from the Context Memory at address $\text{PC}=0$, QGU loads the initial amplitude $\{\alpha^{(\bm{0})}_{j}\}$ from Ping Amplitude Memory and the initial new amplitude $\{\alpha^{(\bm{1})}_{j}\}$ from Pong Amplitude Memory. This data is processed by Gate $0$ in $N$ iterations to produce the updated new amplitude $\{\alpha^{(\bm{0})}_{j}\}$, increment the PC for the next instruction and clear the data in Ping Memory (new initial amplitude is zero). At $\text{PC}=1$, QGU proceeds to read the context of Gate $1$, read amplitude $\{\alpha^{(\bm{1})}_j\}$ from Pong Amplitude Memory, and new amplitude $\{\alpha^{(\bm{2})}_j\}$ from Ping Amplitude Memory. After executing and storing the updated amplitude $\{\alpha^{(\bm{2})}_j\}$ in Ping Amplitude Memory, the PC is incremented by 1, and Pong Amplitude Memory is cleared. This alternating process continues, updating the new amplitude data and storing it in Ping/Pong Amplitude Memory with each gate execution. Throughout FQsun’s computation, the host PS continuously reads and checks the Done flag to determine when the hardware has completed its task. Accordingly, the host PS reads the final amplitude $\{\alpha^{(\bm{m})}_j\}$ from Pong or Ping Amplitude Memory if $m$ is odd or even, respectively. Typically, FQsun's working schedule ensures that the host PS runs the FQsun session accurately and successfully. 
	
	\subsection{MULTIPLE NUMBER PRECISIONS}
	
	The use of various numerical precisions significantly impacts accuracy, hardware resources, and execution speed. Unfortunately, almost all related work on quantum emulators has primarily focused on performance evaluation while lacking detailed criteria for accuracy metrics. In this study, FQsun is designed with five configurations, each corresponding to a different precision: 16-bit floating point ($\bf{FP16}$), 32-bit floating point ($\bf{FP32}$), 16-bit fixed point ($\bf{FX16}$), 24-bit fixed point ($\bf{FX24}$), and 32-bit fixed point ($\bf{FX32}$). A comparison of the architectures, number of pipeline stages, range, and complexity is shown in Table~\ref{tab:precision_comp}. The following discussion elaborates on the key considerations:
	
	\textbf{Pipeline and Complexity}: FP numbers utilize two pipeline stages for both the multiplier and the adder/subtractor units, while FX numbers require only two stages for the multiplier and one stage for the adder/subtractor. FP operation modules are more complex due to handling exponent, mantissa, normalization, and rounding operations. In contrast, FX operations are simpler and involve integer-based arithmetic with fixed multipliers, adders, or subtractors. As a result, the latency of FX design is lower compared to FP design.
	
	\textbf{Range and Precision}: following normalize condition, $|\alpha_j^{(\bm{t})}|\in[-1, 1]$; then, FX numbers used in FQsun are configured with $1$ sign bit, $1$ integer bit, and the remaining bits allocated to the fractional part to optimize precision. Although FP numbers offer a higher maximum precision compared to FX numbers, they come at the cost of increased hardware resources and execution time. It is essential to consider the required accuracy for applications to determine the most appropriate numerical precision.
	
	Overall, FP numbers provide higher accuracy but incur higher latency and resources, while FX numbers offer lower accuracy with reduced hardware resource usage and latency. To identify the most suitable configuration, detailed measurements will be conducted in the following section.

	\begin{table}[]
		\caption{Preliminary analysis of five suitable number precisions for FQsun}
		\label{tab:precision_comp}
		\renewcommand{\arraystretch}{1.2}
		\begin{tabular}{|p{40pt}|p{35pt}|p{50pt}|p{65pt}|}
			\hline
			\textbf{Format} & \textbf{Pipeline stages} & \textbf{Range (Precision)} & \textbf{Complexity} \\ \hline
			FP16 (1-bit sign, 5-bit exponent, 10-bit mantissa) 
			& Two for multiplier, two for adder/subtr. 
			& $[6.1\times10^{-5}, 6.6\times10^{-4}]$ with 10-bit precision 
			& High due to 5-bit exponent, 10-bit mantissa handling \\ \hline
			
			FP32 (1-bit sign, 8-bit exponent, 23-bit mantissa) 
			& Two for multiplier, two for adder/subtr. 
			& $[1.2\times10^{-38}, 3.4\times10^{38}]$ with 23-bit precision 
			& Higher due to 8-bit exponent, 23-bit mantissa handling, normalization, and rounding \\ \hline
			
			FX16 (1-bit sign, 1-bit integer, 14-bit fractional) 
			& Two for multiplier, one for adder/subtr. 
			& $[-1,1]$ with 14-bit precision ($6.10 \times 10^{-5}$) 
			& Lower complexity, only integer-based arithmetic with fixed operators \\ \hline
			
			FX24 (1-bit sign, 1-bit integer, 22-bit fractional) 
			& Two for multiplier, one for adder/subtr. 
			& $[-1,1]$ with 22-bit precision ($2.38 \times 10^{-7}$) 
			& Lower complexity, similar to FX16 but with finer precision due to additional fractional bits \\ \hline
			
			FX32 (1-bit sign, 1-bit integer, 30-bit fractional) 
			& Two for multiplier, one for adder/subtr. 
			& $[-1,1]$ with 30-bit precision ($9.31 \times 10^{-10}$) 
			& Lower complexity, similar to FX24 but with finer precision due to additional fractional bits \\ \hline
			
		\end{tabular}
	\end{table}

	\begin{figure}
		\centering
		\includegraphics[width=\linewidth]{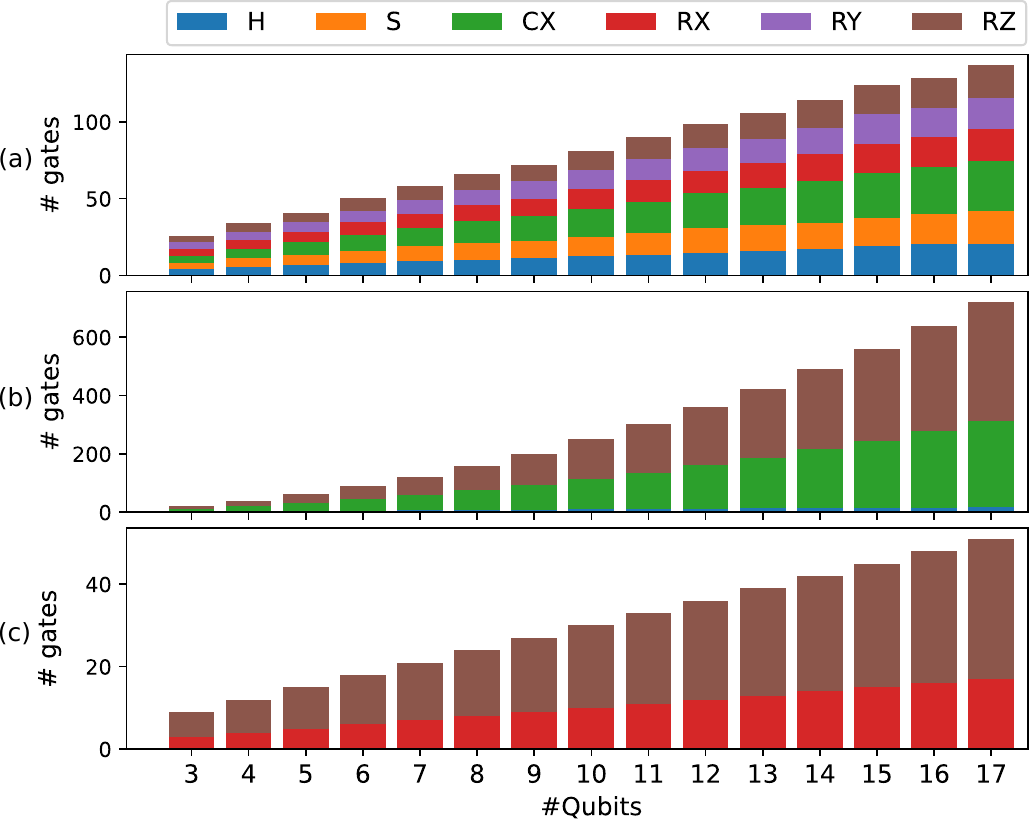}
		\caption{Gate distribution for (a) RQC (b) QFT and (c) PSR.}
		\label{fig:gate}
	\end{figure}

	\section{VERIFICATION AND RESULTS}
	\label{sec:result}
	
	\subsection{TASKS USED FOR BENCHMARKING}
	\label{sec:bencmarking_set}
	
	We chose three tasks to benchmark our proposed emulator with \#Qubits from $3$ to $17$. They are motivated by the quantum machine learning applications \cite{schuld2021machine, Guan_2021} including the Random Quantum Circuits (RQC) sampling \cite{arute2019quantum}, quantum differentiable with Parameter-Shift Rule (PSR) technique \cite{Wierichs2022generalparameter}, and the Quantum Fourier Transform (QFT) as a core algorithmic component to many applications \cite{wakeham2024inferenceinterferenceinvariancequantum}. Since QFT is one of the simple, standard algorithms for benchmarking the performance of quantum simulators and emulators, we test RQC to understand how the proposed universal hardware performs. Finally, PSR is the most popular technique used in the quantum machine learning applications. In Figure~\ref{fig:gate}, we illustrate the gate distribution after transpiling the original circuit by the Clifford+$R_i$ set. The scaling of $\#\text{gates}$ is polynomial complexity based on $\#\text{Qubits}$. The quantum circuit presentation of the PSR and QFT are shown in Figure~\ref{fig:overview} (Inset left) and (Inset right), respectively. Where the circuit sampling from RQC is the combination of used gates with the smallest depth. 
	
	\subsubsection{Random Quantum Circuits (RQC)}
	\label{sec:rqc}
	
	The pseudo-RQC is a popular benchmarking problem for quantum systems, the problem was first implemented on the Sycamore chip to prove quantum supremacy \cite{arute2019quantum}. The quantum circuit is constructed gate-by-gate by randomly choosing from a fixed pool and is designed to minimize the circuit depth. The sequences of gates $\{g_j\}$ create a ``supremacy circuit'' with high classical computational complexity. The RQC dataset from \cite{arute2019quantum} provided the quantum circuit up to $53$ qubits, $1,113$ single-qubit gates, and $430$ two-qubit gates, then proved that the quantum computer could perform this task in seconds instead of a thousand years on classical computers. However, this dataset limits users when they want to benchmark different cases. Based on our previous work \cite{hairivf2024}, we generated a bigger RQC dataset, which used an extendable gate pool and different circuit construction strategies. The size of the generated dataset is $14$ GB of $15,000$ random circuits with a depth range from $1$ to $10$ and $\#$Qubits from $3$ to $17$.

	\subsubsection{Quantum Fourier Transform}
	\label{sec:qft}
	
	Quantum Fourier Transform (QFT) \cite{coppersmith2002approximatefouriertransformuseful} is an important component in many quantum algorithms, such as phase estimation \cite{kitaev1995quantummeasurementsabelianstabilizer}, order-finding, and factoring \cite{365700}. QFT circuit is utilized from Hadamard ($H$), controlled phase ($CP(\theta)$), and SWAP gates where $CP(\theta)$ and SWAP can be decomposed into $R_z(\theta)$ and $CX$ gates. These gates transform on an orthonormal basis $\{|0\rangle,\ldots,|N-1\rangle\}$ with the below action on the basis state $|j\rangle\rightarrow\frac{1}{\sqrt{N}}\sum_{k=0}^{N-1}e^{2\pi ijk/N}$. If the state $|j\rangle$ is written by using the $n$-bit binary representation $j=j_1 ... j_n$:
	
	\begin{equation}
		\begin{split}
			&\text{QFT}(|j_1,\ldots,j_n\rangle)\rightarrow \\&\frac{(|0\rangle+e^{2\pi i 0.j_n}|1\rangle)\otimes\ldots \otimes(|0\rangle+e^{2\pi i 0.j_1 j_2 \ldots j_n}|1\rangle)}{\sqrt{N}}
		\end{split}
	\end{equation}
	
	where $0.j_l\ldots j_n = j_l/2+\ldots+j_n/2^{n-l+1}$. We apply the QFT on the zero state $|\bm{0}\rangle$, which should return the exact equal superposition of all possible $n$-qubit states, means $\alpha_j=\alpha_k \;\forall j,k\in[0,N)$.

	\begin{table}[]
		\centering
		\renewcommand{\arraystretch}{1.2}
		\caption{Properties of benchmarking tasks with different gate types and number of gates.}
		\begin{tabular}{|l|l|l|l|}
			\hline
			\textbf{Task}  & \textbf{RQC}  & \textbf{QFT}  & \textbf{PSR}  \\ \hline
			{Gate}  & Clifford+$R_i$  & $H, CX, R_z$  & $R_x, R_z$  \\ \hline
			{Output}  & $\{|\alpha|^2\}$  & $\{|\alpha|^2\}$  & $\{\bm{\theta}^{(\bm{t+1})}, |\alpha|^2\}$  \\ \hline
			\multirow{3}{*}{$\#\text{Gates}$}  
			& $n \times d$  
			& $n$ ($H$)  
			& $2n$ ($R_z$) \\ 
			&  
			& $(n+3)(n-1)$ ($CX$)  
			& $n$ ($R_x$) \\  
			&  
			& $\frac{3}{2} n(n-1)$ ($R_z$)  
			& \\ \hline
		\end{tabular}
		\label{tab:tasks}
	\end{table}

	\subsubsection{Quantum differentiable programming (QDP) with Parameter-Shift Rule (PSR)}\label{sec:psr}
	
	Starting with $ZXZ(\bm{\theta})$, we try to optimize associated cost value $C(\bm{\theta})$, which is known as the target of the optimization process by first-order optimizers such as Gradient Descent (GD), for example:

	\begin{align}
		C(\bm{\theta}) = \sum_{j=0}^{N-1} j|\alpha_j|^2;\;\bm{\theta}_j^{'}\leftarrow \bm{\theta}_j - \gamma \nabla_{\bm{\theta}} C(\bm{\theta})
		\label{eq:optimizer}
	\end{align}

	where $\bm{\theta}$ is updated by the general PSR technique \cite{Wierichs2022generalparameter}. Because the parameterized gate set is limited as a one-qubit rotation gate, only 2-term PSR is used. The gradient is notated as $\nabla_{\bm{\theta}} C(\bm{\theta}) = \{\partial_{\theta_j} C(\bm{\theta})\}_{j=0}^{m-1}$ and partial derivative $\partial_{\theta_j} C(\bm{\theta})$ is presented in Equation~\eqref{eq:psr} with $\bm e_j$ is the $j^{\text{th}}$-unit vector:
	
	\begin{align}
		\partial_{\theta_j} C(\bm{\theta}) = \frac{1}{\sqrt{2}}[C(\bm{\theta} + \frac{\pi}{2} \bm e_j) - C(\bm{\theta} - \frac{\pi}{2} \bm e_j)]
		\label{eq:psr}
	\end{align}
	
	The optimal parameter $\bm{\theta}^*$ can be obtained through $2m \times n_{\text{iter}}$ quantum evaluations after $n_{\text{iter}}$ iterations.
	
	\subsection{COMPARISON METRICS}
	\label{sec:metrics}
	
	Essentially, using the proposed simulators without knowledge about their metrics, such as error rate and execution time, can lead to bias during the experiment. Therefore, we propose two types of metrics: accuracy for reliability and performance for comparison between FQsun and other emulators/simulators.
	
	\textbf{Accuracy metrics}: As mentioned in Section~\ref{sec:background}, the target of the quantum emulator is to simulate the transition function $\mathcal{W}(g_j)$, receives $|\psi^{(\bm{t-1})}\rangle$ and returns $|\psi^{(\bm{t})}\rangle$. To measure the similarity between computational states $\rho$ and theoretical state $\sigma$, trace fidelity $\mathcal{F}$
	is defined as $\mathcal{F}(\rho, \sigma)=\left(\text{Tr}(\sqrt{\sqrt{\rho} \sigma \sqrt{\rho}})\right)^2$ where the theoretical state $\sigma$ is presented in double precision. Since computing the square root of a positive semi-definite matrix consumes a lot of resources, the fidelity can be reduced to $\left|\left\langle\psi_\rho \mid \psi_\sigma\right\rangle\right|^2$ for $\rho=\left|\psi_\rho\right\rangle\left\langle\psi_\rho\right|$ and $\sigma=\left|\psi_\sigma\right\rangle\left\langle\psi_\sigma\right|$. $\mathcal{F}(\rho, \sigma)= 1$ means $\rho\equiv\sigma$. 
	
			\begin{table}[]
		\centering
		\caption{Utilization of FQSun on ZCU102 FPGA}
		\label{tab:utilization}
		\renewcommand{\arraystretch}{1.2}
		\begin{tabular}{|p{20pt}|p{75pt}|p{22pt}|p{15pt}|p{22pt}|p{15pt}|}
			\hline
			\textbf{Design} & \textbf{Name} & \textbf{LUTs} & \textbf{FFs} & \textbf{BRAM} & \textbf{DSP} \\ \hline
			
			$\bf{FP16}$ & Buffers \& AXI Mapper & 268  & 290  & 0   & 0  \\ \cline{2-6} 
			& Context Memory & 4,428 & 0    & 0   & 0  \\ \cline{2-6} 
			& Ping \& Pong Memories & 3,071 & 36   & 464 & 0  \\ \cline{2-6} 
			& QGU & 5,530 & 2,752 & 0   & 44 \\ \cline{2-6} 
			& FQsun Controller & 937  & 973  & 0   & 0  \\ \cline{2-6} 
			& \textbf{Total} & \textbf{14,234} & \textbf{4,051} & \textbf{464} & \textbf{44} \\ \hline
			
			$\bf{FP32}$ & Buffers \& AXI Mapper & 295  & 327  & 0   & 0  \\ \cline{2-6} 
			& Context Memory & 6,556 & 0    & 0   & 0  \\ \cline{2-6} 
			& Ping \& Pong Memories & 3,796 & 36   & 456 & 0  \\ \cline{2-6} 
			& QGU & 6,258 & 5,296 & 0   & 88 \\ \cline{2-6} 
			& FQsun Controller & 1,188 & 1,372 & 0   & 0  \\ \cline{2-6} 
			& \textbf{Total} & \textbf{18,093} & \textbf{7,031} & \textbf{456} & \textbf{88} \\ \hline
			
			$\bf{FX16}$ & Buffers \& AXI Mapper & 264  & 289  & 0   & 0  \\ \cline{2-6} 
			& Context Memory & 4,387 & 0    & 0   & 0  \\ \cline{2-6} 
			& Ping \& Pong Memories & 3,168 & 36   & 464 & 0  \\ \cline{2-6} 
			& QGU & 457  & 142   & 0   & 14 \\ \cline{2-6} 
			& FQsun Controller & 990  & 973  & 0   & 0  \\ \cline{2-6} 
			& \textbf{Total} & \textbf{9,266} & \textbf{1,440} & \textbf{464} & \textbf{14} \\ \hline
			
			$\bf{FX24}$ & Buffers \& AXI Mapper & 272  & 307  & 0   & 0  \\ \cline{2-6} 
			& Context Memory & 5,722 & 0    & 0   & 0  \\ \cline{2-6} 
			& Ping \& Pong Memories & 3,086 & 24   & 344 & 0  \\ \cline{2-6} 
			& QGU & 684  & 206   & 0   & 28 \\ \cline{2-6} 
			& FQsun Controller & 939  & 828  & 0   & 0  \\ \cline{2-6} 
			& \textbf{Total} & \textbf{10,703} & \textbf{1,635} & \textbf{344} & \textbf{28} \\ \hline
			
			$\bf{FX32}$ & Buffers \& AXI Mapper & 297  & 327  & 0   & 0  \\ \cline{2-6} 
			& Context Memory & 6,888 & 0    & 0   & 0  \\ \cline{2-6} 
			& Ping \& Pong Memories & 3,934 & 36   & 456 & 0  \\ \cline{2-6} 
			& QGU & 914  & 326   & 0   & 56 \\ \cline{2-6} 
			& FQsun Controller & 1,136 & 1,372 & 0   & 0  \\ \cline{2-6} 
			& \textbf{Total} & \textbf{13,169} & \textbf{2,061} & \textbf{456} & \textbf{56} \\ \hline
			
		\end{tabular}
	\end{table}
	
	Since there are a limited number of decimals on the quantum emulator, it makes the systematic error on amplitudes $\{\alpha_j\}$ increase as system size. Because $|\{\alpha_j\}|=2^n$ and $\sum_j |\alpha_j|^2 = 1$, then $\mathbb{E}[\alpha_j]=1/2^n$ decrease exponentially based on \#Qubits. $\text{MSE}(|\psi_{\rho}\rangle, |\psi_{\sigma}\rangle)$ is calculated as $\frac{1}{2^n}\left(\sum_j|\alpha_j^{\rho} - \alpha_j^{\sigma}|^2\right)$ to get the accumulated error of all amplitudes.
	
	\textbf{Performance metrics}: The novelty of FQsun is the high performance, measured via execution time, power-delay product (PDP), and memory efficiency. The execution time ($t$) is counted by creating a quantum circuit to receive amplitudes. PDP is simply equal to Power (Joules) $\times\;t$. When the designed quantum emulator consumes extra-low power compared with CPU/GPU, this metric is important for verifying the performance of FQsun.
	
	\begin{figure*}
		\centering
		\includegraphics[width=0.99\linewidth]{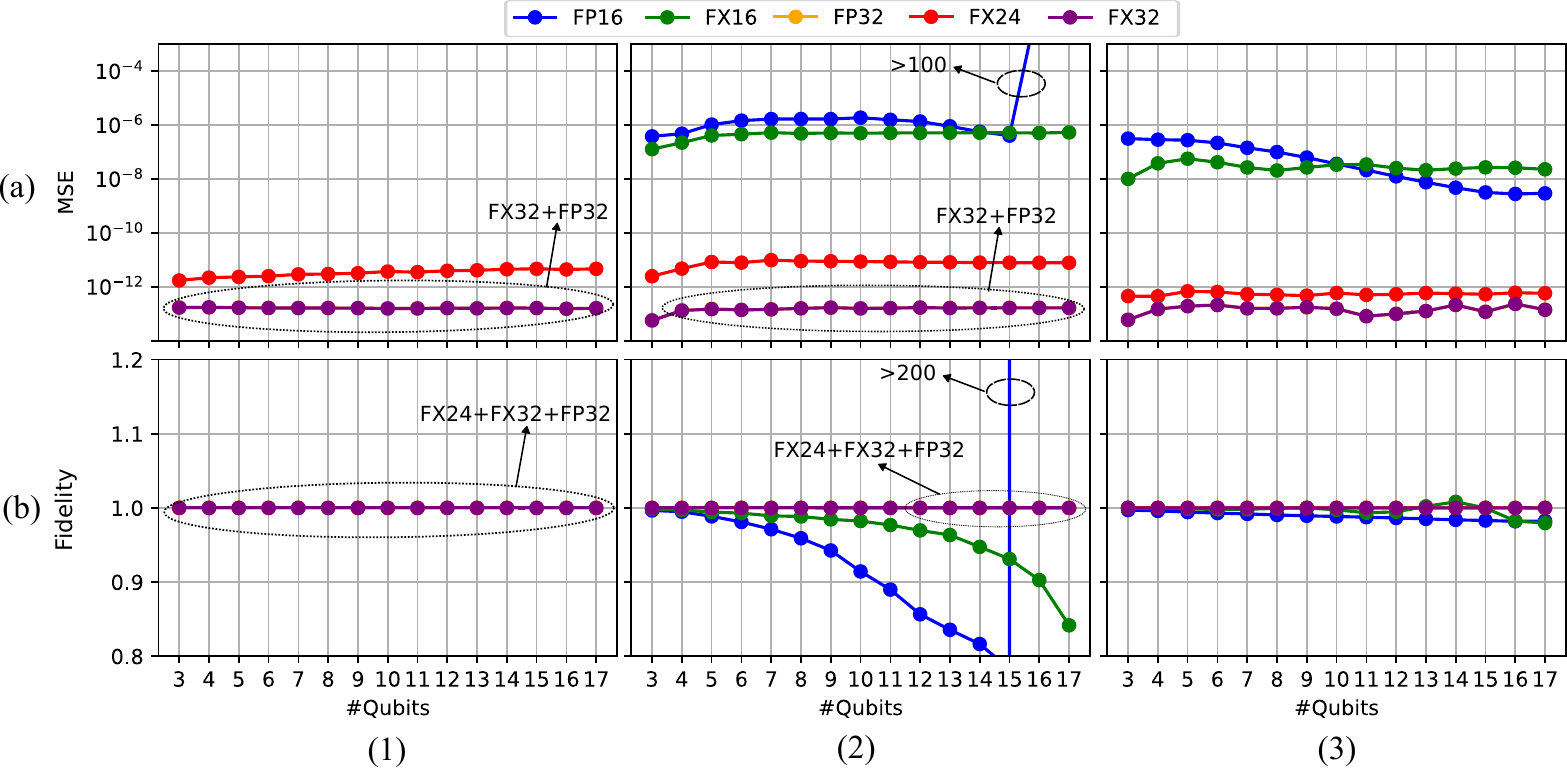}
		\caption{(a) $\text{MSE}_{\text{$\bf{FQsun}$}}$ and (b) $\mathcal{F}_{\text{$\bf{FQsun}$}^*}$. All three tasks include (1) RQC (at $d=10$), (2) QFT, and (c) PSR. In all cases, $\bf{FP32}$ and $\bf{FX32}$ are overlapped. For QFT, the $\bf{FP16}$ version performs the wrong results since it can not present any $\alpha_j\leq1/2^{16}\approx10^{-5}$.}
		\label{fig:qft_psr_mse}
	\end{figure*}

	\subsection{IMPLEMENTATION AND RESOURCE UTILIZATION ON FPGA}
	\label{sec:fpga}
	
	The FQsun designs were implemented using Verilog HDL and realized on a 16nm Xilinx ZCU102 FPGA, utilizing Vivado 2021.2 to obtain synthesis and implementation results. The system underwent testing through five designs with different precision: $\bf{FP16}$, $\bf{FX16}$, $\bf{FX24}$, $\bf{FP32}$, $\bf{FX32}$. For convenience in presentation, we denote the following sets: $\bf{Software}$ = $\{$PennyLane, Qiskit, ProjectQ, Qsun$\}$, $\bf{FQsun} = \{\bf{FP16}, \bf{FX16}, \bf{FX24}, \bf{FP32},  \bf{FX32}\}$ and $\bf{FQsun}^{*} = \bf{FX24}, \bf{FP32}, \bf{FX32}\}$.

	\begin{table}[]
		\centering
		\caption{Basic gate's period and number of cycles per gate on $\bf{FQsun}$}
		\renewcommand{\arraystretch}{1.3}
		\label{tab:gate_time}
		\begin{tabular}{|c|c|cccccc|}
			\hline
			\multirow{2}{*}{\textbf{Version}} &
			\multirow{2}{*}{\textbf{Period (ns)}} &
			\multicolumn{6}{c|}{\textbf{Number of Cycles per Gate}} \\ \cline{3-8} 
			&
			&
			\multicolumn{1}{c|}{$H$} &
			\multicolumn{1}{c|}{$S$} &
			\multicolumn{1}{c|}{$CX$} &
			\multicolumn{1}{c|}{$R_x$} &
			\multicolumn{1}{c|}{$R_y$} &
			$R_z$ \\ \hline
			$\bf{FP16}$ & $6.66$ & \multicolumn{1}{c|}{$6$} & \multicolumn{1}{c|}{$4$} & \multicolumn{1}{c|}{$4$} & \multicolumn{1}{c|}{$6$} & \multicolumn{1}{c|}{$6$} & $8$ \\ \hline
			$\bf{FP32}$ & $7.35$ & \multicolumn{1}{c|}{$6$} & \multicolumn{1}{c|}{$4$} & \multicolumn{1}{c|}{$4$} & \multicolumn{1}{c|}{$6$} & \multicolumn{1}{c|}{$6$} & $8$ \\ \hline
			$\bf{FX16}$ & $7.35$ & \multicolumn{1}{c|}{$4$} & \multicolumn{1}{c|}{$2$} & \multicolumn{1}{c|}{$2$} & \multicolumn{1}{c|}{$4$} & \multicolumn{1}{c|}{$4$} & $4$ \\ \hline
			$\bf{FX24}$ & $8.00$ & \multicolumn{1}{c|}{$4$} & \multicolumn{1}{c|}{$2$} & \multicolumn{1}{c|}{$2$} & \multicolumn{1}{c|}{$4$} & \multicolumn{1}{c|}{$4$} & $4$ \\ \hline
			$\bf{FX32}$ & $9.35$ & \multicolumn{1}{c|}{$4$} & \multicolumn{1}{c|}{$2$} & \multicolumn{1}{c|}{$2$} & \multicolumn{1}{c|}{$4$} & \multicolumn{1}{c|}{$4$} & $4$ \\ \hline
		\end{tabular}
	\end{table}

	Subsequently, three primary applications as described in Section~\ref{sec:bencmarking_set} were implemented to demonstrate computational efficiency and high flexibility. System accuracy across each design will be recognized by considering fidelity $\mathcal{F}$ and MSE. Consequently, the efficiency and practicality of each FQsun quantum emulator design will be clearly illustrated, alongside detailed presentations of quantum simulator packages on powerful CPUs. Table~\ref{tab:utilization} provides a detailed report on the utilization of $\bf{FQsun}$ when implemented on the Xilinx ZCU102, based on the number of lookup tables (LUTs), flip-flops (FFs), Block RAMs (BRAMs), and digital signature processors (DSPs). Accordingly, the designs utilize between 9,226 and 18,093 LUTs, 1,440 to 7,031 FFs, 344 to 464 BRAMs, and 14 to 88 DSPs. It should be noted that the $\bf{FP16}$ and $\bf{FX16}$ versions exhibit the highest BRAM utilization due to their capacity to accommodate up to 18 qubits. Conversely, the remaining configurations ($\bf{FQsun}$$^*$) support maximally $17$ qubits, resulting in a reduced BRAM requirement. The significantly lower DSP utilization compared to FP numbers demonstrates that applying FX numbers to the FQsun emulator results in greater power efficiency due to each DSP being equivalent to multiple LUTs in terms of implementation.

	Overall, $\bf{FX32}$ utilizes the most LUTs, while $\bf{FP32}$ utilizes the most FFs and DSPs. $\bf{FX16}$ and $\bf{FP16}$ utilize the fewest resources. Additionally, all FQsun versions are designed to avoid utilizing DSPs to conserve hardware resources.

	\subsection{GATE SPEED}
	\label{sec:gate-speed}

	The properties of gate speed are presented in Table~\ref{tab:gate_time}. The execution time per gate is given by $\#\text{Period}_{\text{version}} \times \#\text{Cycle} \times 2^n$ for looping through amplitudes, which depends on the version, gate type, and number of qubits, respectively. These results highlight the trade-off between precision and speed across different versions. For example, while the $\bf{FP16}$ version has the shortest period of \(6.66\) ns, the $\bf{FX32}$ version requires a longer period of \(9.35\) ns to execute a gate, reflecting the additional time needed to achieve higher precision. Furthermore, the FX versions generally require fewer cycles per gate compared to FP versions, with reductions ranging from $1.5$ to $2$ times depending on the gate type. For example, the $CX$ gate requires $4$ cycles in $\bf{FX16}$ but 6 cycles in $\bf{FP16}$, a difference that becomes critical in high-speed applications. This efficiency in FX versions could be advantageous in time-sensitive quantum computations, where lower \#Cycles allow for faster gate operations while maintaining adequate precision. Therefore, depending on the application, a balance between speed (fewer \#Cycles) and accuracy (higher precision) can be strategically chosen to optimize performance.

	\subsection{MEAN-SQUARE ERROR EVALUATION}
	
	\begin{figure}
		\centering
		\includegraphics[width=1\linewidth]{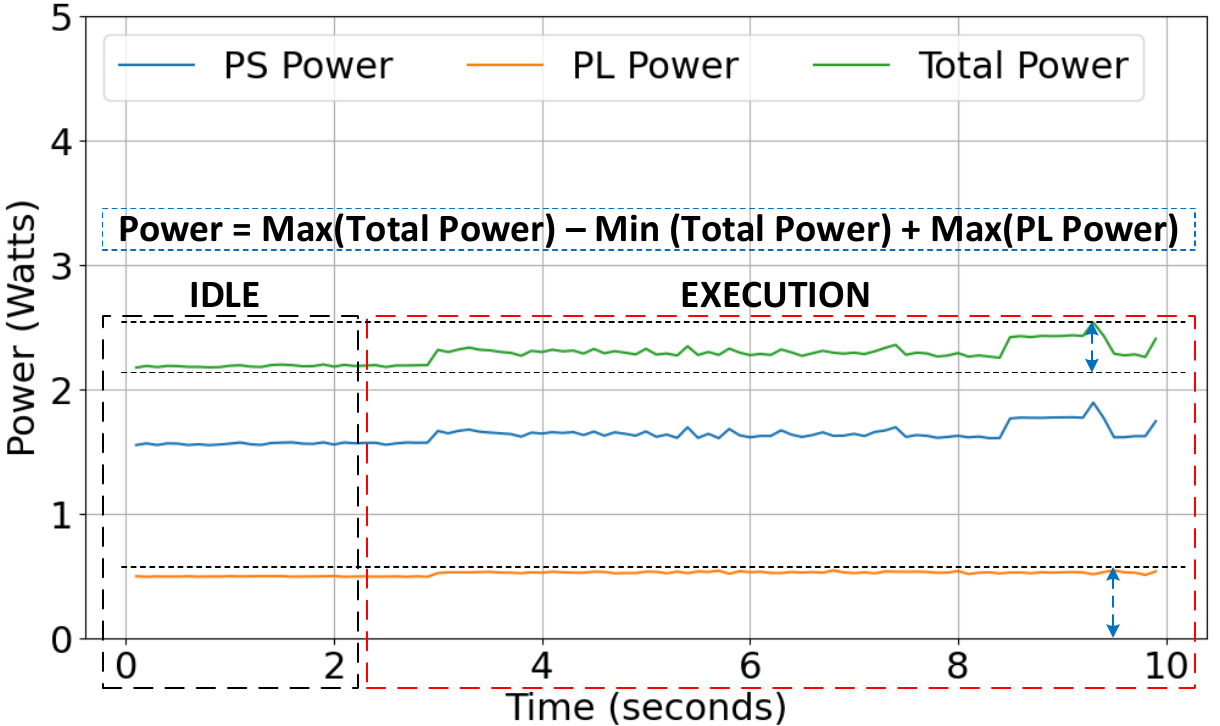}
		\vspace{-7mm}
		\caption{Power measurement of FQsun ($\bf{FX32}$) on RQC.}
		\vspace{-7mm}
		\label{fig:power_measure}
	\end{figure}
	
	\begin{figure*}
		\centering
		\includegraphics[width=0.99\linewidth]{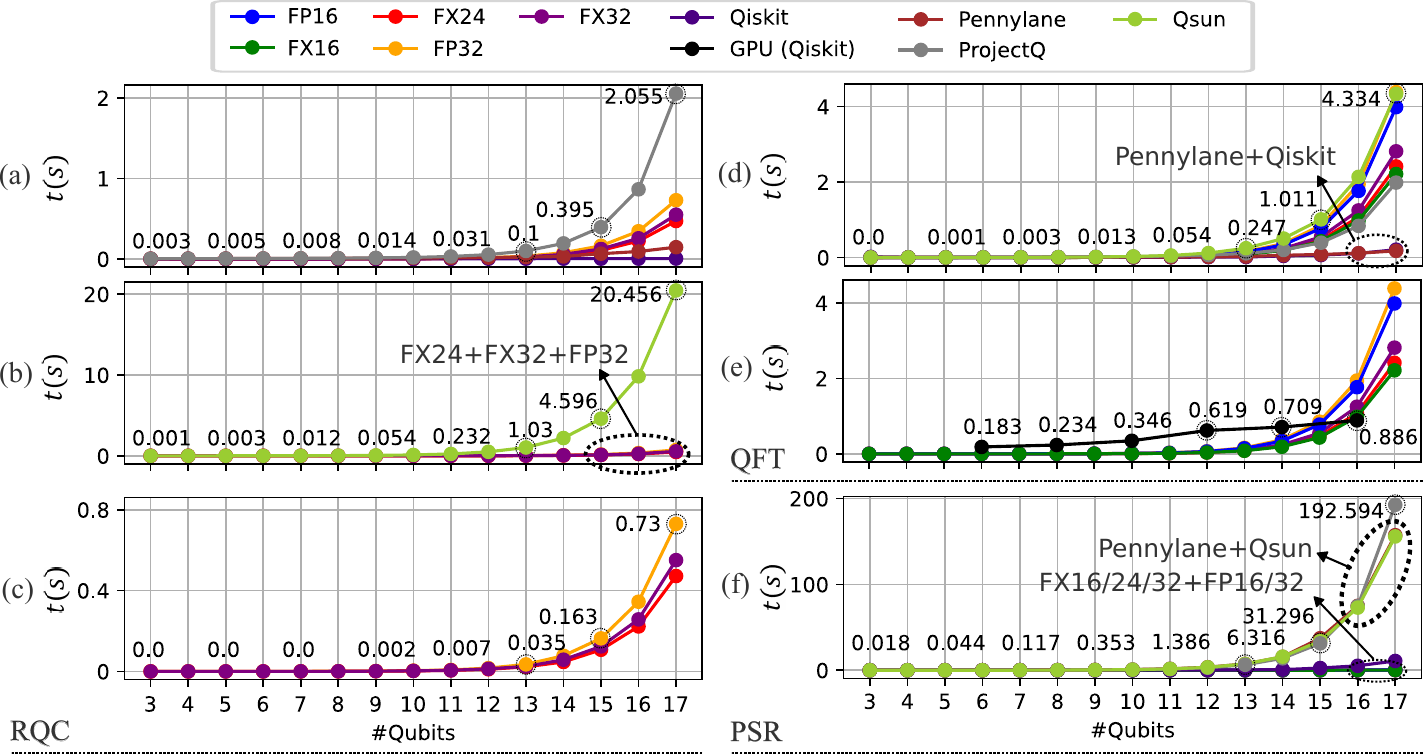}
		\caption{The execution time on (a-c) RQC, compare between (a) $\bf{Software}$/$\{\text{Qsun}\}$ + $\bf{FQsun}$$^*$, (b) Qsun + $\text{$\bf{FQsun}$}^*$ and (c) $\text{$\bf{FQsun}$}^*$. $d=10$. (d) QFT execution time comparison between $\bf{Software}$ + $\bf{FQsun}$. (e) We compare GPU A100 \cite{jamadagni2024benchmarkingquantumcomputersimulation} and $\bf{FQsun}$. Since the A100 data is only available on even \#Qubits, we only plot until $n=16$. FQsun demonstrates a clear speed advantage for $n<15$. (f) PSR execution time comparison between $\bf{Software}$ + $\bf{FQsun}$. The figures on the plot from (a) to (f) are from ProjectQ, Qsun, FP32, Qsun, GPU (Qiskit,) and ProjectQ, respectively.}
		\label{fig:qrc_exe_time}
	\end{figure*}

	To demonstrate the accuracy, the measured MSE of $\bf{FQsun}$ on the real-time ZCU102 FPGA system is shown in Figure \ref{fig:qft_psr_mse}. Based on the analyzed data, the designs $\bf{FQsun}$$^*$ exhibit the lowest MSE values, making them the most suitable candidates for use in quantum emulation. Specifically, $\bf{FP32}$ and $\bf{FX32}$ demonstrate extremely low MSE values ranging from $5.686 \times 10^{-14}$ to $1.656 \times 10^{-13}$, indicating greatly high accuracy. $\bf{FX24}$ also shows acceptable MSE values, ranging from $2.481 \times 10^{-12}$ to $7.798 \times 10^{-12}$, which are sufficient for large quantum applications. In contrast, $\bf{FP16}$ and $\bf{FX16}$ do not meet the required accuracy standards. $\bf{FP16}$ displays significantly high MSE values, with MSE reaching $0.055$ at $n = 16$ and spiking to $129.824$ at $n = 17$, rendering it unsuitable for precise quantum applications. Similarly, $\bf{FX16}$ shows higher MSE values, ranging from $1.278 \times 10^{-7}$ to $5.327 \times 10^{-7}$, which are inadequate for applications demanding high precision. Thus, the three designs in $\bf{FQsun}$$^*$ with the lowest MSE: $\bf{FX24}$, $\bf{FP32}$, and $\bf{FX32}$, will be considered for comparison in the next sections.
	
	\subsection{FIDELITY EVALUATION}

	While MSE can provide a preliminary assessment of accuracy in quantum emulators, fidelity is the more precise metric for evaluating the performance of quantum simulation systems. The results for RQC, as shown in Figure \ref{fig:qft_psr_mse} (1), emphasize the accuracy gap between the three designs. Specifically, $\mathcal{F}_{\bf{FP32}/\bf{FX32}(\text{RQC})}\in[1 - 7.6\times 10^{-7},1 + 1.2\times10^{-7}]$, remaining measurably close to $1$. Both designs demonstrate superior stability and precision, making them the best performers among the three. By contrast, $\mathcal{F}_{\bf{FX24}(\text{RQC})}\in[1 - 7.6\times 10^{-7}, 1 - 2\times 10^{-6}]$ shows a broader range of fidelity. For QFT and PSR in Figure \ref{fig:qft_psr_mse} (2-3), demonstrate again that $\mathcal{F}_{\bf{FP32}/\bf{FX32}(\text{QFT}/\text{PSR})}\in[1-7\times10^{-6},1 + 2.4\times10^{-5}]$, have prominently small variations. Both designs exhibit stable performance, with values consistently close to ideal accuracy. This positions $\bf{FP32}$ and $\bf{FX32}$ as the optimal choices in terms of fidelity. On the other hand, $\mathcal{F}_{\bf{FX24}(\text{QFT}/\text{PSR})}\in[1 - 6\times10^{-5}, 1 + 3\times 10^{-5}]$,  still shows slightly larger variations. While $\bf{FX24}$ maintains a high level of accuracy, these more noticeable deviations suggest a lower stability compared to $\bf{FP32}$ and $\bf{FX32}$.

	To maintain the accuracy of FPGA, we list all the feasible data structures. We also evaluate the impact of data structure on the above tasks. Consistent with the commentary from \cite{10653682}, there is a trade-off between high fidelity and quantization bit-width. As the small scale, the required bit-width for $\mathcal{F}\approx 1$ is low but increases fast based on $\#\text{Qubits}$ \cite{10.1063/1.3497087}.

	\subsection{DETAILED POWER CONSUMPTION ANALYSIS}

	To accurately measure power consumption in real time, the INA226 sensor on the ZCU102 FPGA was utilized. Detailed results are shown in Figure \ref{fig:power_measure}, where the power consumption values for the PS, PL, and total power are presented when $\bf{FP32}$ executes RQC. Specifically, the PS consumes a maximum of $1.84$ W (with $0.34$ W dynamic power). The PL consumes a maximum of $0.61$ W. The total power reaches a maximum of $2.41$ W (with $0.23$ W dynamic power). Accordingly, the power consumption for FQsun is determined by adding the total dynamic power to the maximum PL power, reaching $0.81$ W. These power values are kept for all $\#$Qubits.

	\subsection{COMPARISON WITH RELATED SOFTWARE ON POWERFUL CPU/GPU}

	To demonstrate the speed advantage of FQsun, the simulation time of $\bf{FQsun}$$^*$ is presented and compared with $\bf{Software}$. All software trials are run on an Intel i9-10940X CPU @ 3.30GHz at least $100$ times then take the average.

	The comparisons in Figure \ref{fig:qrc_exe_time} (a-c) are conducted with PSR on different set versions. In Figure~\ref{fig:qrc_exe_time} (a), $\bf{FQsun}$ shows comparable execution speed with most software simulators at $n<13$. When compared with Qsun, $\bf{FQsun}^{*}$ shows faster computational speed, especially for $n>11$, demonstrating a significant improvement over the corresponding software version. Finally, a comparison between $\bf{FQsun}^{*}$ versions reveals that the FX version provides faster computational speed, with the difference increasing gradually based on $\#$Qubits. Overall, FQsun offers better processing speed than ProjectQ and Qsun emulators and is slower than Qiskit and PennyLane. At the same time, FX versions are considered to have the best processing speed. Depending on the accuracy requirements, an appropriate version can be selected.
	
	Figure \ref{fig:qrc_exe_time} (d-e) shows the mean of the execution time when simulating QFT and PSR. In the QFT benchmark, $\bf{FQsun}$ is slower than Qiskit and PennyLane for $n\in[11,17]$. However, it demonstrates superior speed compared to other software simulations and is slightly faster than Qiskit. For larger \#Qubits, the GPU's performance is slightly better due to its parallel processing property. The highlight of FQsun is its significantly higher energy efficiency, consuming less than $1$ W of power, while GPUs typically consume hundreds. In Figure~\ref{fig:qrc_exe_time} (f), Qiskit still achieves top performance but is lower than all $\bf{FQsun}$ versions, while Pennylane and Qsun are overlapped; ProjectQ got the worst performance with 192.598 (s) at $n=17$.

	\begin{figure}
		\centering
		\includegraphics[width=1\linewidth]{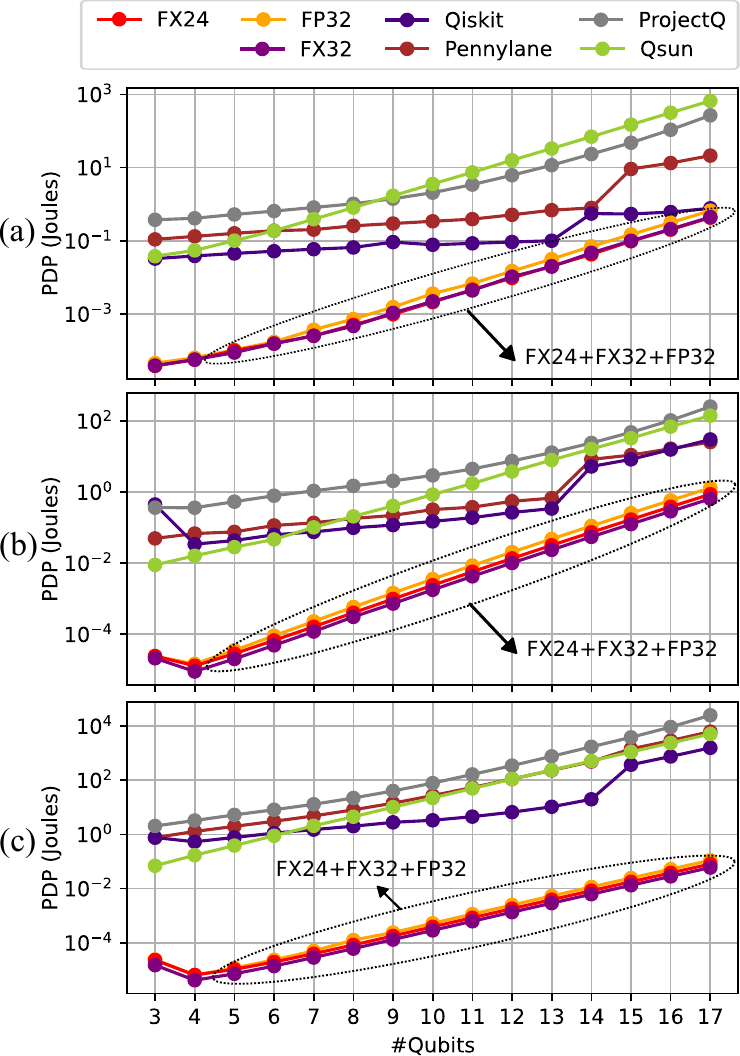}
		\caption{PDP on $\bf{Software}$ + $\bf{FQsun}^*$ through (a) RQC (b) QFT and (c) PSR. The PDP comparison results show that FQsun has optimal energy efficiency, making it suitable for the long term to save costs. }
		\label{fig:pdp}
	\end{figure}

	To further demonstrate energy efficiency, Figure \ref{fig:pdp} presents a comparison of PDP between $\bf{FQsun}^{*}$ and four software simulations. Accordingly, FQsun achieves significantly better PDP than software simulations in all three tasks. In QFT, FQsun provides a PDP from $1.30\times 10^{-5}$ to $1.31$, better than $\bf{Software}$ from $\bm{3.62\times10^2}$ to $\bm{4.0\times 10^4}$ times; and from $4.14\times 10^{-6}$ to $8.04\times 10^{-2}$, better than $\bf{Software}$ from $\bm{3.18\times 10^{3}}$ to $\bm{7.84\times 10^{5}}$ times when simulating PSR. Finally, for RQC, FQsun achieves from $2.56\times 10^{-5}$ to $3.72\times 10^{-1}$, better than $\bf{Software}$ from $\bm{1.66\times 10^{0}}$ to $\bm{9.87\times 10^{3}}$ times.

	\subsection{COMPARISON WITH QISKIT BASED ON STABILIZER ORDER}
	\label{sec:compare_stabilizer_order}

	The results from the previous section show that Pennylane and Qiskit can achieve better execution times than $\bf{FQsun}$ in RQC and QFT tasks from $n>10$. Pennylane and Qiskit are the multi-paradigm simulation packages, hence, it will use the best paradigm for each type of circuit rather than the full state-vector as ProjectQ, Qsun, and $\bf{FQsun}$. Therefore, Pennylane and Qiskit will provide different gate speeds for different circuits. The paradigm is decided based on the stabilizer order property, which ranges from $1$ to $4^n$. The stabilizer order relates to the complexity of the circuit; the more non-Clifford on qubits, the higher the stabilizer order. Using stabilizer formalism, we can simulate low-order circuits like QFT in only $0.11$ (s) at $n=64$ \cite{modelcounting}, much better than the state-vector approach because QFT uses non-Clifford gates on only the last qubits. In the case of PSR, the high-order circuits use many non-Clifford gates, which require the state-vector simulation, and then Pennylane and Qiskit are slower than FQsun. 
	
\begin{figure}
		\centering
		\includegraphics[width=1\linewidth]{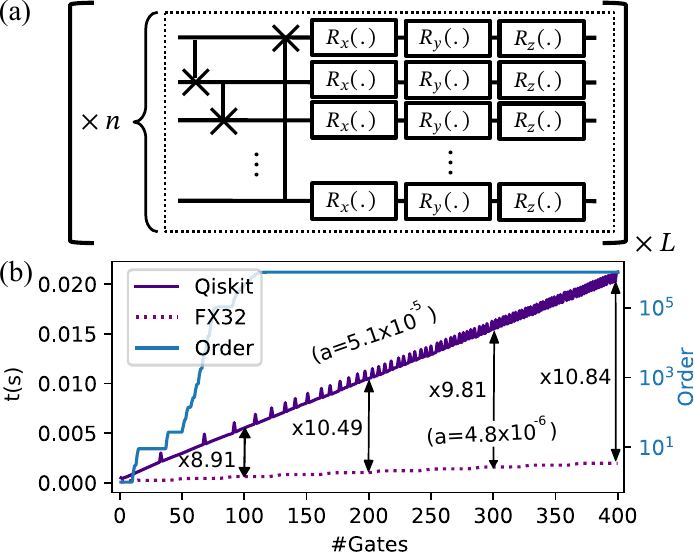}
		\caption{(a) 10-layer $|W_{\text{chain}}+XYZ(\bm\theta)\rangle$ is used for benchmarking in Section~\ref{sec:compare_stabilizer_order} (b) The execution time between Qiskit and $\bf{FX32}$ based on the stabilizer order (in logarithm scale).}
		\label{fig:order}
	\end{figure}

	For a fair comparison, we evaluate Qiskit and $\bf{FX32}$ on $10$ - qubits $W_{\text{chain}}+XYZ$ ansatz as shown in Figure~\ref{fig:order} (a) to make sure that all simulators follow the state-vector paradigm. Figure~\ref{fig:order} (b) demonstrates how the gap between Qiskit and our emulator changed based on the stabilizer order. Execution times are measured again for each additional gate. The alternating action of CNOT and non-Clifford gates increases the order based on $\#$Gates, until $4^{10}$ for all stabilizers. Notice that the number of stabilizers is equal to $\#$Qubits. The results show that $\bf{FX32}$ achieve the slope better than Qiskit $\textbf{10.71}$ times, take only $1.96\times10^{-3}$ (s) compared to $2.1\times10^{-2}$ (s) at $\#\text{Gate}=400$. It can be explained that in the case of $W_{\text{chain}}+ZXZ$, the stabilizer is quickly increased to maximum between $\#\text{Layers}=2$ ($\#\text{Gates}=80$) and $\#\text{Layers}=3$ ($\#\text{Gates}=120$).

	\subsection{COMPARISON WITH HARDWARE-BASED QUANTUM EMULATORS}

	Table \ref{tab:qft} presents a comparison between FQsun and existing FPGA-based quantum emulators \cite{10653682, Mahmud2020EfficientCT, 9952408, 8115369, 9798809} in terms of frequency, precision, execution time,$\#$Gates, and normalized gate speed. $\#$Qubits stand for the maximum of supported $\#$Qubits. All figures are taken on their own FPGA. 
	
	Compared to other quantum emulators \cite{10265215, 10653682,10.1109/ISVLSI.2008.43,1347938, https://doi.org/10.1155/2016/5718124}, our emulator supports computation with a higher \#Qubits ($17$ compared to under $5$), demonstrating enhanced applicability for various applications. The work  \cite{10.1109/ISVLSI.2008.43, 1347938, https://doi.org/10.1155/2016/5718124} achieve the best normalized gate speed from $10^{-10}$ to $10^{-9}$ (s). However, the hardware design is fixed for $\#\text{Qubits}$ such as $3$; then, it only takes one cycle per gate application. Furthermore, the reported execution time is measured on simulation and does not implement SoC. The work \cite{10653682} and \cite{9798809} also show a lower execution time than our work. Again, these designs support only the QFT circuit and then have no use case because they provide a fixed output. 
	
	For a fair comparison, FQsun is comparable to other flexible emulators that can potentially support a wide range of quantum algorithms, including Arria 10AX115N4F45E3SG \cite{Mahmud2020EfficientCT} and AMD Xilinx Zynq-7000 FPGA \cite{8115369}. Regarding execution time, $\bf{FX32}$ achieves a computation speed-up of $\textbf{17.9}$ times and $\textbf{6.1}$ times compared to \cite{Mahmud2020EfficientCT} and \cite{8115369}, respectively. Furthermore, FQsun supports computations across five precision types and offers different performance and accuracy.
	
	\section{CONCLUSION}
	\label{sec:conclusion}
	
	In conclusion, the FQsun demonstrates significant advancements in quantum emulators, delivering improved speed, accuracy, and energy efficiency compared to traditional software simulators and existing hardware emulators. Leveraging optimized memory architecture, a configurable QGU, and support for multiple number precisions, FQsun efficiently bridges the gap between simulation flexibility and hardware constraints. Experimental results reveal that FQsun achieves high fidelity and minimal MSE across various quantum tasks, outperforming comparable emulators on execution time and PDP. The emulator's flexibility in supporting diverse quantum gates enables versatile applications, positioning FQsun as a reliable platform for advanced quantum simulation research. 
	
	Our proposed hardware proves that a quantum emulator can be more efficient than the same type that runs on CPU/GPU. To improve the performance of the emulator, the accelerated technique on software can be implemented on hardware. Further acceleration techniques, such as gate-fusion \cite{10.1145/3126908.3126947}, accelerated PSR \cite{hai2024}, and realized-state representation \cite{https://doi.org/10.4218/etrij.2021-0442} will be considered.

	\begin{table*}[ht]
		\centering
		\renewcommand{\arraystretch}{1.2}
		\resizebox{0.99\linewidth}{!}{\begin{threeparttable}
				\caption{Comparative analysis of FQsun and existing FPGA-based emulators on QFT's performance.}
				\label{tab:qft}
				\begin{tabular}{|c|c|c|c|c|c|c|c|c|}
					\hline
					\textbf{Works} &
					\textbf{Device} & \textbf{Frequency (MHz)}
					& \textbf{Flexibility} &
					\textbf{Precision} &
					\textbf{\#Qubits} &
					\begin{tabular}[c]{@{}c@{}}\textbf{Execution}  \textbf{time (s)}\end{tabular} &  
					\textbf{\#Gates $^{\dagger}$} &
					\textbf{NGS $^{\dagger\dagger}$} \\ \hline
					
					\cite{10653682} &
					Xilinx XCVU9P &
					233 & \XSolidBrush &
					18-bit FX &
					16 & 
					$1.20\times 10^{-3}$ & -
					& -
					\\ \hline
					
					\cite{Mahmud2020EfficientCT} &
					\begin{tabular}[c]{@{}c@{}}Arria\\ 10AX115N4F45E3SG\end{tabular} &
					233 & \Checkmark &
					32-bit FP &
					16 & 
					$1.84\times10^{1}$ & 528
					& $5.33\times 10^{-7}$
					\\ \hline
					
					\cite{9952408} &
					\begin{tabular}[c]{@{}c@{}}Xilinx XCKU115 \end{tabular} &
					160 & \XSolidBrush &
					16-bit FX &
					16 &
					$2.70\times 10^{-1}$ & 136
					& $3.03\times 10^{-8}$
					\\ \hline

					\cite{8115369} &
					\begin{tabular}[c]{@{}c@{}}AMD Xilinx\\ Zynq-7000 \end{tabular} &
					100 & \Checkmark &
					32-bit FX &
					6 & 
					$1.15\times 10^{-4}$ & 10
					& $1.80\times 10^{-7}$
					\\ \hline
					
					\cite{9798809}&
					\begin{tabular}[c]{@{}c@{}}2 $\times$ Intel Stratix 10\\ MX2100 \end{tabular} &
					299 & \XSolidBrush &
					32-bit FP &
					30 &
					$4.47\times 10^0$ & 465
					& $8.95\times 10^{-12}$
					\\ \hline
					\multirow{5}{*}{This work} &
					\multirow{5}{*}{Xilinx ZCU102} &
					150 & \Checkmark &
					16-bit FP &
					18 &
					$8.90\times 10^0$ & 810
					& $4.19\times 10^{-8}$
					\\ \cline{3-9} 
					&
					&
					136 & \Checkmark &
					32-bit FP &
					17 &
					$4.30\times 10^0$ & 721
					& $4.55\times 10^{-8}$
					\\ \cline{3-9} 
					&
					&
					136 & \Checkmark &
					16-bit FX &
					18 &
					$4.90\times 10^0$ & 810
					& $2.31\times 10^{-8}$
					\\ \cline{3-9} 
					&
					&
					125 & \Checkmark &
					24-bit FX &
					17 &
					$2.41\times 10^0$ & 721
					& $2.52\times 10^{-8}$
					\\ \cline{3-9} 
					&
					&
					125 & \Checkmark &
					32-bit FX &
					17 &
					$2.81 \times 10^0$ & 721
					& $2.97\times 10^{-8}$
					\\ \hline
				\end{tabular}
				\begin{tablenotes}
					\item[$\dagger$] The $\#\text{Gate}$ in this work is higher than other work due to the no-use of control-rotation gates.
					\item[$\dagger\dagger$] The \textbf{N}ormalized \textbf{G}ate \textbf{S}peed (NGS) (s / (gate $\times$ amplitude)) = Execution time / (\#Gates $\times$ $2^{\#\text{Qubits}}$), smaller is better.
				\end{tablenotes}
		\end{threeparttable}}
	\end{table*}

	\section*{Data availability}
	
	The codes and data used for this study are available at https://github.com/NAIST-Archlab/FQsun.
	
	\section*{Acknowledgment}
	
	This work was supported by JST-ALCA-Next Program Grant Number JPMJAN23F4, Japan, and partly executed in response to the support of JSPS, KAKENHI Grant No. 22H00515, Japan.
	
	\bibliographystyle{IEEEtran}
	\bibliography{ref.bib}

	\begin{IEEEbiography}[{\includegraphics[width=1in,height=1.25in,clip,keepaspectratio]{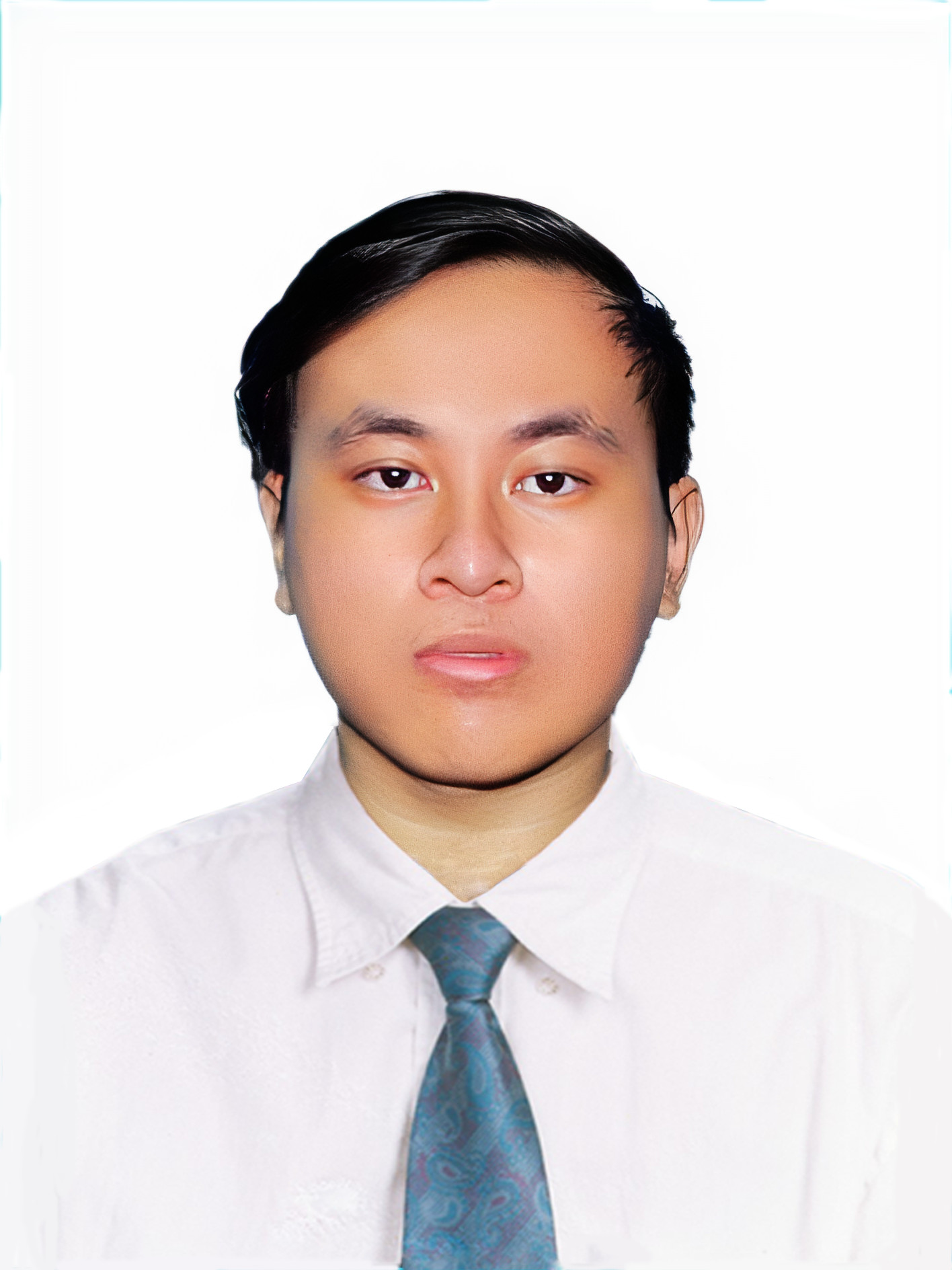}}]{TUAN HAI VU } received the B.S. degree in software engineering and M.S. degree in computer science from the University of Information Technology, Vietnam National University, in 2021 and 2023, respectively. He is currently a Ph.D. student at the Architecture Lab at Nara Institute of Science and Technology, Japan, from 2024. His research interests include quantum simulation acceleration and quantum machine learning.
	\end{IEEEbiography}
	
	\vspace{-15mm}
	
	\begin{IEEEbiography}[{\includegraphics[width=1in,height=1.25in,clip,keepaspectratio]{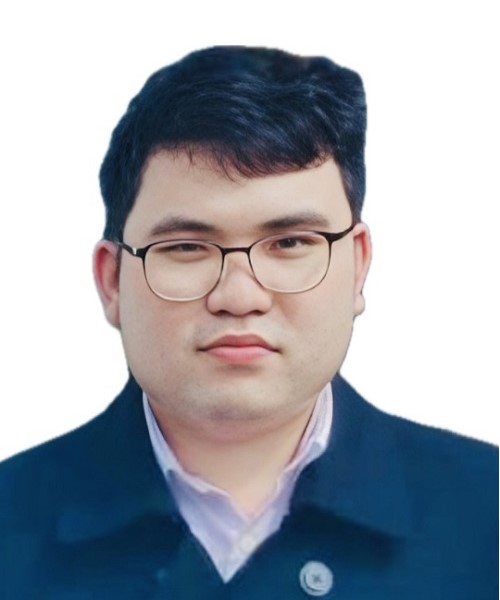}}]{VU TRUNG DUONG LE } received the Bachelor of Engineering degree in IC and hardware design from Vietnam National University Ho Chi Minh City (VNU-HCM)—University of Information Technology (UIT) in 2020, and the Master’s degree in information science from the Nara Institute of Science and Technology (NAIST), Japan, in 2022. He completed his Ph.D. degree in 2024 at NAIST, where he now serves as an Assistant Professor at the Computing Architecture Laboratory. His research interests include computing architecture, reconfigurable processors, and accelerator design for quantum emulators and cryptography.      
	\end{IEEEbiography}
	
	\vspace{-15mm}
	
	\begin{IEEEbiography}[{\includegraphics[width=1in,height=1.25in,clip,keepaspectratio]{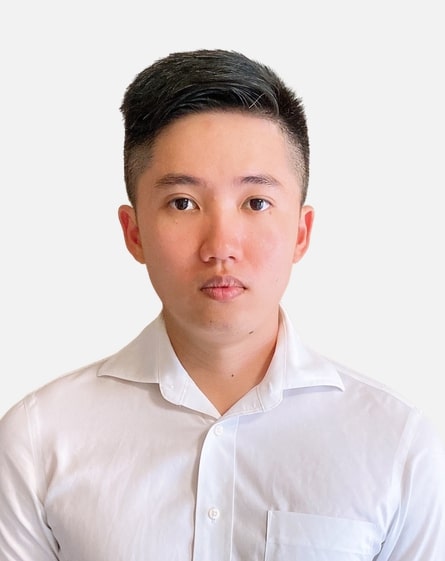}}]{HOAI LUAN PHAM } received a bachelor’s degree in computer engineering from Vietnam National University Ho Chi Minh City—University of Information Technology (UIT), Vietnam, in 2018, and a master’s degree and Ph.D. degree in information science from the Nara Institute of Science and Technology (NAIST), Japan, in 2020 and 2022, respectively. Since October 2022, he has been with NAIST as an Assistant Professor and with UIT as a Visiting Lecture. His research interests include blockchain technology, cryptography, computer architecture, circuit design, and accelerators.      
	\end{IEEEbiography}
	
	
	\vspace{-15mm}
	
	\begin{IEEEbiography}[{\includegraphics[width=1in,height=1.25in,clip,keepaspectratio]{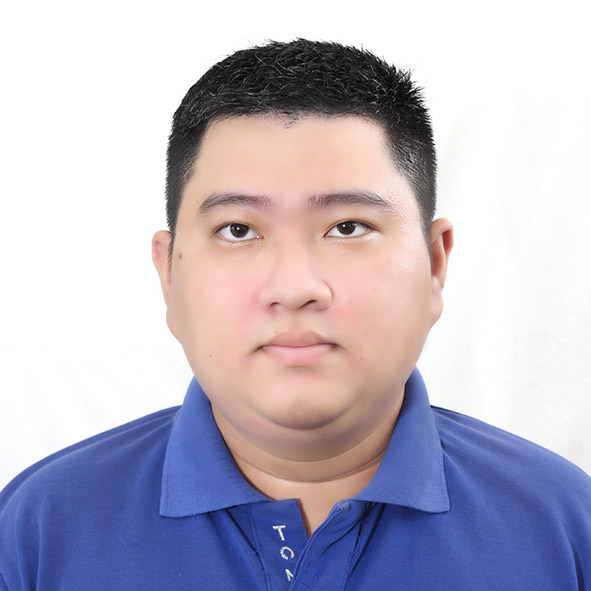}}]{QUOC CHUONG NGUYEN } received the B.S. degree in theoretical physics from the University of Natural Sciences, Vietnam National University, in 2018. He is currently a Ph.D. student at the State University of New York at Buffalo, the United States of America, starting in 2023. His research interests include quantum machine learning and algorithms for discrete optimization problems.      
	\end{IEEEbiography}
	
	\vspace{-15mm}
	
	\begin{IEEEbiography}[{\includegraphics[width=1in,height=1.25in,clip,keepaspectratio]{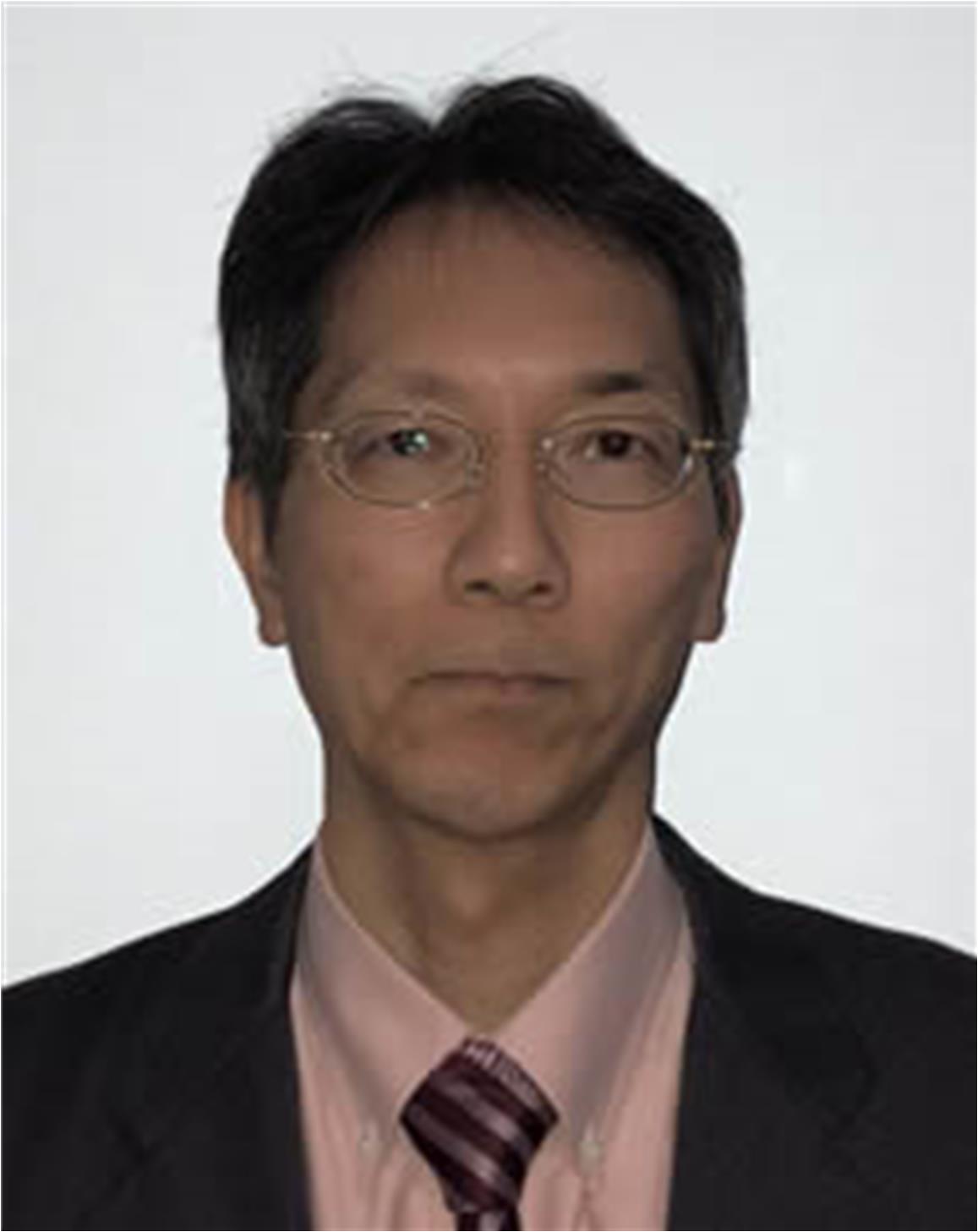}}]{YASUHIKO NAKASHIMA } received B.E., M.E., and Ph.D. degrees in Computer Engineering from Kyoto University in 1986, 1988, and 1998, respectively. He was a computer architect in the Computer and System Architecture Department, FUJITSU Limited, from 1988 to 1999. From 1999 to 2005, he was an associate professor at the Graduate School of Economics, Kyoto University. Since 2006, he has been a professor at the Graduate School of Information Science, Nara Institute of Science and Technology. His research interests include computer architecture, emulation, circuit design, and accelerators. He is a fellow of IEICE, a senior member of IPSJ, and a member of IEEE CS and ACM.
	\end{IEEEbiography}

	\EOD
	
\end{document}